\documentclass[
reprint,
amsmath,amssymb, aps, prd,
]{revtex4-2}
\usepackage[T1]{fontenc}
\usepackage{lmodern}
\usepackage[utf8]{inputenc}
\usepackage{tikz}
\usetikzlibrary{decorations.markings, arrows.meta}
\graphicspath{{figures/}}
\usepackage{graphicx}
\usepackage{dcolumn}
\usepackage{bm}
\usepackage{hyperref}
\usepackage{orcidlink}

\begin{document}

\preprint{APS/123-QED}

\title{Axionic Casimir Torque}

\author{Soham Ghatak \orcidlink{0009-0008-6197-8842}}
 \email{soham22@iiserb.ac.in}
\affiliation{
 Department of Physics,\\ Indian Institute of Science Education and Research, Bhopal\\ Madhya Pradesh---462 066, India
}

\date{\today}

\begin{abstract}
In the canonical Casimir effect, the quantum vacuum is isotropic, and the minimization of vacuum energy drives only a linear attractive force between the plates. However, the introduction of an axion gradient breaks this rotational symmetry. It can be demonstrated that these asymmetric zero-point fluctuations exert a macroscopic mechanical twist on the plates proportional to $\vec{b}^2 \sin (2\vartheta)$ where $\vec{b}$ is the spatial axion gradient and $\vartheta$ is the angle between the axion gradient and the normal to the plates.
\end{abstract}

\maketitle

\section{\label{sec:introduction}Introduction}

The Peccei--Quinn mechanism \cite{PecceiQuinn1977} is the most economical resolution of the strong-CP problem. Promoting the CP-violating QCD vacuum angle to a dynamical field $\theta(x)$ automatically relaxes it to zero, at the cost of predicting a new, naturally light pseudoscalar, the axion \cite{Weinberg1978,Wilczek1978}. Beyond solving strong CP, the axion is among the most compelling dark matter candidates in the literature \cite{PreskillWiseWilczek1983,AbbottSikivie1983,DineFischler1983}. This dual role has motivated extensive experiments searching for axion-photon interactions \cite{SikiviePRL1983}. The coupling responsible for essentially all of this phenomenology is the anomalous term $\mathcal{L}_{a\gamma\gamma} = -\frac{1}{4}\theta(x) F_{\mu\nu}\tilde{F}^{\mu\nu}$ analyzed in detail by Wilczek \cite{Wilczek1987}, who showed that a spacetime-dependent $\theta(x)$ modifies Maxwell's equations through effective, axion-induced charge and current densities sourced by $\nabla\theta$ and $\dot\theta$.

The Casimir effect \cite{Casimir1948,KlimchitskayaMostepanenkoMohideen2009} is one place to look for the imprint of this term. In the standard, axion-free case, the $\mathrm{U}(1)$ vacuum is rotationally invariant, and the zero-point energy between two conducting plates can only produce a force along the line joining them. Once $\theta(x)$ is allowed to vary, this isotropy is explicitly broken, and the Casimir problem changes character. Torque itself is not a new observable in Casimir physics: dielectrically anisotropic slabs have long been predicted to experience a fluctuation-induced torque that drives their optical axes into alignment \cite{ParsegianWeiss1972,Barash1978,vanEnk1995,Munday2005,Munday2008Erratum}, an effect later measured directly between a birefringent crystal and a liquid-crystal layer \cite{Somers2018}. What has no precedent in that literature is a torque sourced not by material birefringence but by the axion background itself.

Ref. \cite{Fukushima2019} showed that an axion gradient along the normal to the plate splits the photon dispersion relation into two branches, producing a Casimir force that becomes repulsive for $bL_z \gtrsim 2.38$, bypassing the `no-go theorem' of Ref. \cite{KennethKlich2006}, in a manner analogous to the chiral-medium mechanism of Ref. \cite{JiangWilczek2019}, where an optically active medium is inserted between perfect conductors. Ema et al. \cite{Ema2023} showed that for metal plates coated with Weyl semimetals whose node separation is normal to the plates, there is a stable separation at which the Casimir force vanishes. Ref. \cite{Favitta2023} used Green's functions to compute the Casimir force between perfectly conducting plates to second order in a constant $\dot\theta$, the spectral energy density for an oscillating $\theta(t)=\theta_0\sin\omega_a t$, and the temperature-dependent radiation pressure on a static axion domain wall $\theta(z)$. The constant timelike case has been treated by Ref. \cite{KharlanovZhukovsky2010}, and Ref. \cite{Brevik2021} considered a $z$-dependent axion between plates in a scalar-electrodynamics model. This present work deals with the Casimir torque for two parallel conducting plates in the presence of an axion gradient at an arbitrary angle $\vartheta$ to the plate normal, for both purely spatial and asymptotic spatio-temporal backgrounds. Because the misaligned gradient breaks the residual rotational symmetry about the plate normal, the vacuum energy depends explicitly on $\vartheta$, and its variation defines an observable that has no counterpart in
Refs. \cite{Fukushima2019,Ema2023,Favitta2023}: a Casimir \textit{torque}.

The following \textit{conventions} have been adopted: the use of natural units ($\hbar = c = k_B = 1$), $\eta_{\mu\nu} = \text{diag} (+1, -1, -1, -1)$; all four-vectors are in bold-face ($\mathbf{k}$) and all three-vectors have an overhead arrow ($\vec{k}$); Euclidean vectors are denoted with a subscript `E', and the imaginary unit is written as $\mathbf{i} = \sqrt{-1}$ to distinguish it from the spatial index $i$.

\section{The Effective Lagrangian}

We begin with axion electrodynamics as formulated by Ref. \cite{Fukushima2019}:
\begin{equation} \label{eq:1}
    \mathcal{L} = -\frac{1}{4}F_{\mu\nu}F^{\mu\nu} - \frac{1}{4}\theta F_{\mu\nu}\tilde{F}^{\mu\nu} + \frac{1}{2\xi}(\partial_\mu A^\mu)^2 + \frac{1}{2} \partial_\mu \overline{c} \, \partial^\mu c,
\end{equation}
where the axion coupling term $\theta(x)$ makes the pseudoscalar contraction $F_{\mu\nu}\tilde{F}^{\mu\nu}=- 4\vec E\cdot\vec B$ dynamical, and the gauge-fixing term $\frac{1}{2\xi}(\partial_\mu A^\mu)^2$ renders the otherwise singular Maxwell operator invertible \cite{Fukushima2019}. Writing $\mathcal{L}_\text{photon}=-\frac{1}{2}A_\nu(G^{-1})^{\mu\nu}A_\mu$ and passing to momentum space, the inverse propagator is
\begin{equation} \label{eq:2}
(\tilde{G}^{-1})^{\mu\nu}(\mathbf{k}) = \mathbf{k}^2 \eta^{\mu\nu} - \left(1-\frac{1}{\xi}\right)k^\mu k^\nu + \mathbf{i}\epsilon^{\mu\nu\alpha\beta}b_\alpha k_\beta ,
\end{equation}
a result that holds for arbitrary constant $b_\mu=\partial_\mu\theta$, since no particular direction for $b_\mu$ was assumed in its derivation. We take $\mathbf{b} = (b_0,\, b\sin\vartheta\cos\varphi,\, b\sin\vartheta\sin\varphi,\, b\cos\vartheta)$ where $\vartheta$ is the angle between $\vec b$ and the $z$-axis and $\varphi$ its azimuthal angle in the transverse plane. This Chern--Simons-like structure in Eq. \eqref{eq:2} is the Lorentz- and CPT-violating modification of electrodynamics considered independently by Ref. \cite{CarrollFieldJackiw1990}, with $b_\mu$ there playing the role of a fixed background vector rather than an axion gradient. Diagonalizing Eq. \eqref{eq:2} in the Feynman gauge ($\xi=1$) gives two unphysical eigenvalues $\lambda_1=\lambda_2=\mathbf{k}^2$, canceled by the ghost sector \cite{Fukushima2019}, and two physical eigenvalues $\lambda_\pm=\mathbf{k}^2\pm\sqrt{(\mathbf{b}\cdot\mathbf{k})^2-\mathbf{b}^2\mathbf{k}^2}$ corresponding to the two photon polarizations. Combining the photon and ghost partition functions, the vacuum energy density is
\begin{equation} \label{eq:3}
\mathbf{i} \ln Z = VT\mathcal{E} = -\frac{\mathbf{i}}{2}\sum_{\mathbf{k}}\ln\left[\mathbf{k}^4 + \mathbf{b}^2 \mathbf{k}^2 - (\mathbf{b}\cdot \mathbf{k})^2\right] .
\end{equation}  
This is where we deviate from Ref. \cite{Fukushima2019}.

\section{\label{spatialaxion}A Spatial Axion}

In vacuum, we have the canonical longitudinal momentum quantization $k_3 = n\pi/L_z$ \cite{Casimir1948}. However, a nonzero axion gradient renders the vacuum optically active (the two circular polarizations propagate differently) \cite{Fukushima2019,Favitta2023,CarrollFieldJackiw1990}. For $\vec b$ normal to the plates, perfect conductor boundaries do not couple the two branches: the $\pm$ modes are independent solutions and $k_3=n\pi/L_z$ for both \cite{Ema2023,Fukushima2019}. For a misaligned gradient this decoupling is not guaranteed, and a full treatment of any boundary-induced polarization mixing would require a Lifshitz-type scattering-matrix formalism \cite{Lifshitz1956}, of the kind used for a chiral medium between perfect conductors \cite{JiangWilczek2019}, in the helicity-basis reflection-matrix form of Ref. \cite{Fukushima2019}, and for Weyl-semimetal-coated plates \cite{Ema2023}.

We use an approximation in the same spirit as Ref. \cite{Fukushima2019}, imposing the quantization condition $k_3=n\pi/L_z$ for arbitrary $\vartheta$. For $\vartheta=0$ this is the exact result for perfect conductors \cite{Fukushima2019,Ema2023}; for $\vartheta\neq0$ it is equivalent to treating the two branches as independent fields confined within $L_z$ and neglecting boundary-induced mixing.

We first restrict the axion gradient to a purely spatial configuration, $b^{\mu} = (0, \vec{b})$, where $\vec{b}$ lies in the transverse-longitudinal plane at an angle $\vartheta$ relative to the normal vector. 
\begin{equation} \label{eq:4}
\mathcal{E} = -\frac{\mathbf{i}}{2}\frac{1}{L_z}\sum_{n=0}^{\infty}\int \frac{ \mathrm{d}k_1\, \mathrm{d}k_2}{(2\pi)^2}\int_{-\infty}^{\infty}\frac{\mathrm{d}k_0}{2\pi}\ln \left[ \mathcal{P}(\mathbf{k}) \right] .
\end{equation}
where:
\begin{equation} \label{eq:5}
\mathcal{P} (\mathbf{k}) \equiv \mathbf{k}^4 + \mathbf{b}^2 \mathbf{k}^2 - (\mathbf{b}\cdot \mathbf{k})^2.
\end{equation}
Setting $\mathcal{P}(\mathbf{k}) = 0$ yields the modified dispersion relations for the physical photon modes:
\begin{equation} \label{eq:6}
\omega_\pm^2 = \vec{k}^2 + \frac{\vec{b}^2}{2} \pm \frac{\sqrt{\vec{b}^4  + 4(\vec{b} \cdot \vec{k})^2}}{2}.
\end{equation}

\subsection{\label{energy1}Spatial Energy}

Using Eq. \eqref{eq:5}, $\ln \left[ \mathcal{P} (\mathbf{k}) \right] = \ln (k_0^2 - \omega_+^2) + \ln (k_0^2 - \omega_-^2)$. Therefore, Eq. \eqref{eq:4} now becomes:
\begin{equation} \label{eq:7}
\mathcal{E} = -\frac{\mathbf{i}}{2}\frac{1}{L_z} \sum_\pm \sum_{n=0}^{\infty}\int \frac{ \mathrm{d}k_1\, \mathrm{d}k_2}{(2\pi)^2} I_\pm (\omega_\pm), 
\end{equation}
where the integral, evaluated using a semicircular contour closed in the right-half complex plane, yields:
\begin{equation} \label{eq:8}
I_\pm (\omega_\pm) \equiv \lim_{\epsilon \to 0} \int_{-\infty}^{\infty}\frac{\mathrm{d}k_0}{2\pi}\ln \left[ k_0^2 - \omega_\pm^2 + \mathbf{i} \epsilon \right] = \mathbf{i} \omega_\pm.
\end{equation}
Using the argument principle and then performing a Wick rotation to Euclidean momentum, we find: 
\begin{align}
\omega_+ + \omega_- &= \int_{-\infty}^{\infty}\frac{\mathrm{d}k_4}{2\pi}\ln\mathcal{P}(\mathbf{i}k_4) \notag \\
&= \int_{-\infty}^\infty\frac{\mathrm{d}k_4}{2\pi}\ln\left[(k_4^2+\omega_+^2)(k_4^2+\omega_-^2)\right] .\label{eq:9}
\end{align}
Substituting this result into Eq. \eqref{eq:8} and inserting it into Eq. \eqref{eq:7}:
\begin{eqnarray} \label{eq:10}
\mathcal{E}_\text{plates} &=& \frac{1}{4L_z}\sum_{n=-\infty}^{\infty} \int\frac{\mathrm{d}k_4 \, \mathrm{d}k_1\, \mathrm{d}k_2}{(2\pi)^3} \nonumber \\
& &\times \ln\left[\mathbf{k}_E^4 + \mathbf{b}_E^2 \mathbf{k}_E^2 - (\mathbf{b}_E \cdot \mathbf{k}_E)^2\right].
\end{eqnarray}
Now, Eq. \eqref{eq:10} contains the formally infinite zero-point energy of the vacuum. Using Poisson's summation, we separate this into a continuous bulk limit, representing the unobservable vacuum energy of infinite space, $\mathcal{E}_\text{vacuum}$, while the remaining sum captures the finite-size boundary corrections induced by the plate separation $L_z$. That is, $ \mathcal{E}_\text{plates} = \mathcal{E}_\text{Casimir} + \mathcal{E}_\text{vacuum}$.
\begin{equation} \label{eq:11}
\mathcal{E}_\text{Casimir} = \sum_{m=1}^\infty \int\frac{\mathrm{d}\mathbf{k}_E}{(2\pi)^4} \ln\Big[\mathbf{k}_E^4 + \mathbf{b}_E^2 \mathbf{k}_E^2 - (\mathbf{b}_E \cdot \mathbf{k}_E)^2 \Big] \mathrm{e}^{\mathbf{i}Q_mk_3}.
\end{equation}
where $Q_m = 2mL_z$ and $k_3 = n\pi/L_z$. We rotate the transverse coordinates $(k_1, k_2)$ into a new basis $(k_a, k_b)$ aligned with the spatial background vector $\vec{b}$, defining:
\begin{align}
    k_1 &\equiv k_a\cos\vartheta + k_b \sin\vartheta , \text{ and}\label{eq:12}\\
    k_3 &\equiv k_b\cos\vartheta - k_a \sin\vartheta,\label{eq:13}
\end{align} 
such that $k_1^2 + k_3^2 = k_a^2 + k_b^2$. 
Upon rotation and transition to polar coordinates in the transverse momentum subspace, the Casimir energy evaluates to:
\begin{eqnarray}
\mathcal{E}_\text{Casimir}(\vartheta) &=& -\frac{1}{\pi^2} \sum_{m=1}^{\infty} \int_0^{\infty} \mathrm{d}u \, u \left[ \frac{\sin(Q_m u \sin\vartheta)}{Q_m^2 \cos\vartheta \sin\vartheta} \right] \nonumber \\
& &\times \operatorname{Re}\left[e^{-Q_m \cos\vartheta \sqrt{u^2 + \mathbf{i}\vert{}\vec{b}\vert{}u}}\right].\label{eq:14}
\end{eqnarray}

\subsubsection{Recovering the canonical result}

We examine the result in the limit where the axion background $b \rightarrow 0$.

\begin{equation}
\mathcal{E}_\text{Casimir} = -\frac{1}{\pi^2}\sum_{m=1}^{\infty}\int_0^{\infty} \mathrm{d}u\,u\,     \frac{\sin(Q_m u\sin\vartheta)}{Q_m^2\cos\vartheta\sin\vartheta} \mathrm{e}^{-Q_m u\cos\vartheta} .\label{eq:15}
\end{equation}
For any $\alpha_1$, $\alpha_2$, standard integration yields: $ \int_0^{\infty} \mathrm{d}u \, u \sin(\alpha_1 u) \exp[-\alpha_2 u] = \frac{2\alpha_1\alpha_2}{(\alpha_1^2 + \alpha_2^2)^2}$. Let $\alpha_1 = Q_m \sin\vartheta$ and $\alpha_2 = Q_m \cos\vartheta$. Therefore:
\begin{align}
\mathcal{E}_\text{Casimir} &= -\frac{1}{\pi^2} \sum_{m=1}^{\infty} \frac{1}{Q_m^2 \cos\vartheta \sin\vartheta} \left[ \frac{2 \sin\vartheta \cos\vartheta}{Q_m^2} \right] \notag \\
&= -\frac{1}{8\pi^2 L_z^4} \zeta(4) = -\frac{\pi^2}{720 L_z^4} .\label{eq:16}
\end{align}

\subsection{\label{force1}Spatial Force}

The force per unit volume is obtained by taking the derivative of Eq. \eqref{eq:14} with respect to $L_z$ and introducing an overall negative sign, yielding: 
\begin{equation} \label{eq:17}
\mathcal{F}(b) = -\frac{2}{\pi^2} \sum_{m=1}^{\infty} m \int_{0}^{\infty} \mathrm{d}u \, u \, \text{Re} \; [ \mathrm{e}^{-\lambda R_+} ( \tilde{A} R_+ + \tilde{C} )],
\end{equation}
where $\lambda = Q_m \cos\vartheta$ and 
$R_\pm = \sqrt{u^2 \pm \mathbf{i}\vert{}\vec{b}\vert{}u}$ along with the auxiliary functions:
\begin{align}
\tilde{A}(\vartheta, u) &\equiv \frac{\sin(Q_m u \sin\vartheta)}{Q_m^2 \sin\vartheta} , \text{ and}\label{eq:18}\\
\tilde{C}(\vartheta, u) &\equiv \frac{2\sin(Q_m u \sin\vartheta)}{Q_m^3 \cos\vartheta \sin\vartheta} - \frac{u \cos(Q_m u \sin\vartheta)}{Q_m^2 \cos\vartheta} .\label{eq:19}
\end{align}

\begin{figure}
\centering    \includegraphics[width=1\linewidth]{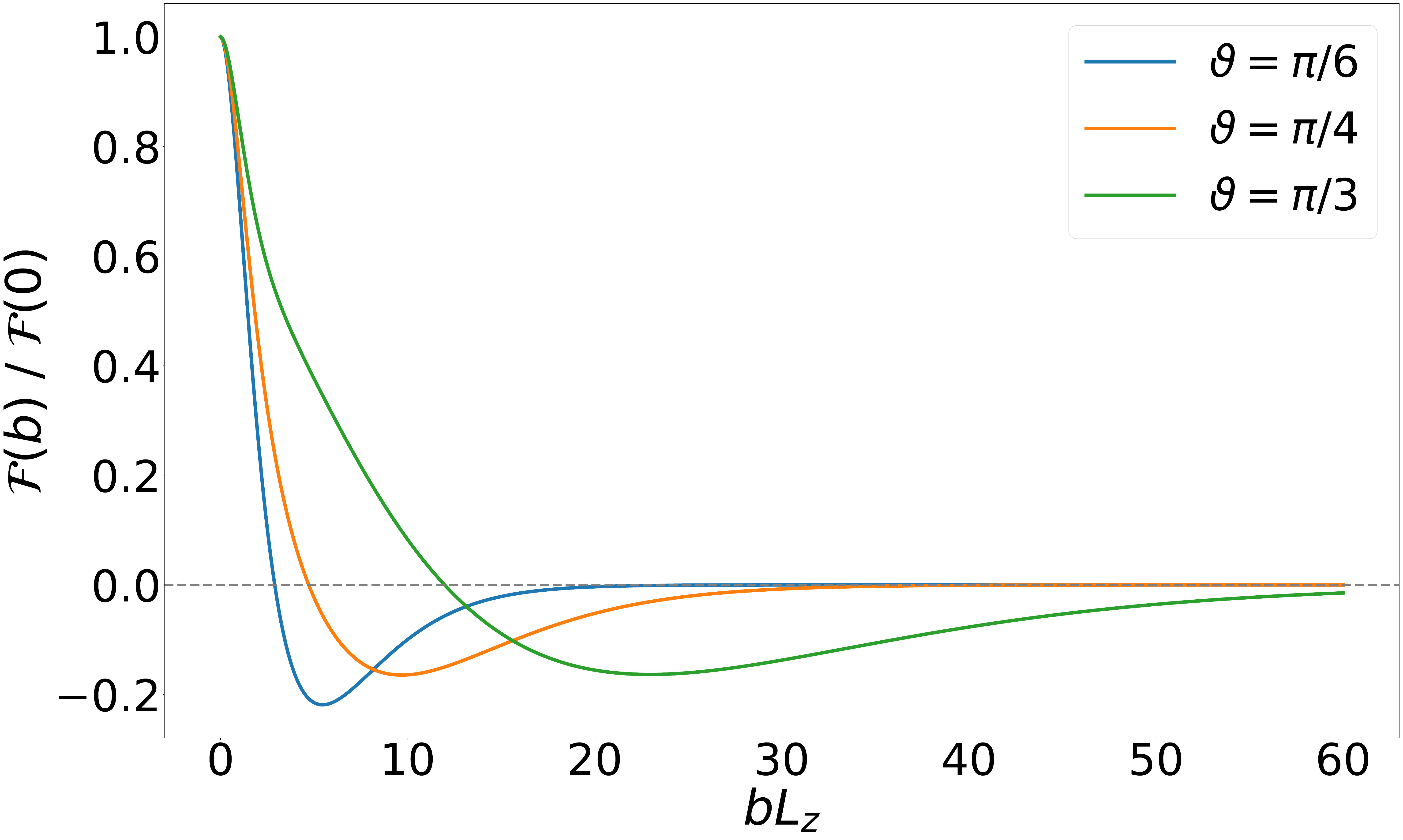}
\caption{The Casimir force per unit volume plotted against $bL_z$ for three different values of $\vartheta$. At the origin where $bL_z = 0$, the normalized ratio $\mathcal{F}(b)/\mathcal{F}(0)$ is 1.0, showing that as $b \to 0$, the formula recovers the canonical Casimir force. At high values of $bL_z$, all three angular curves asymptotically approach zero, ensuring that it does not diverge. Thus the force can be tuned to exactly zero or pushed into a purely repulsive regime where the normalized curve drops below the zero axis. In Appendix \ref{AppendixC} we show that this reduces to the exact force per unit area expression as derived by Ref. \cite{Fukushima2019} when $\vartheta \to 0$.}
\label{fig:F_b__F_0_vs_bLz}
\end{figure}

\subsection{\label{torque1}Spatial Torque}

\begin{figure}
\centering
\includegraphics[width=1\linewidth]{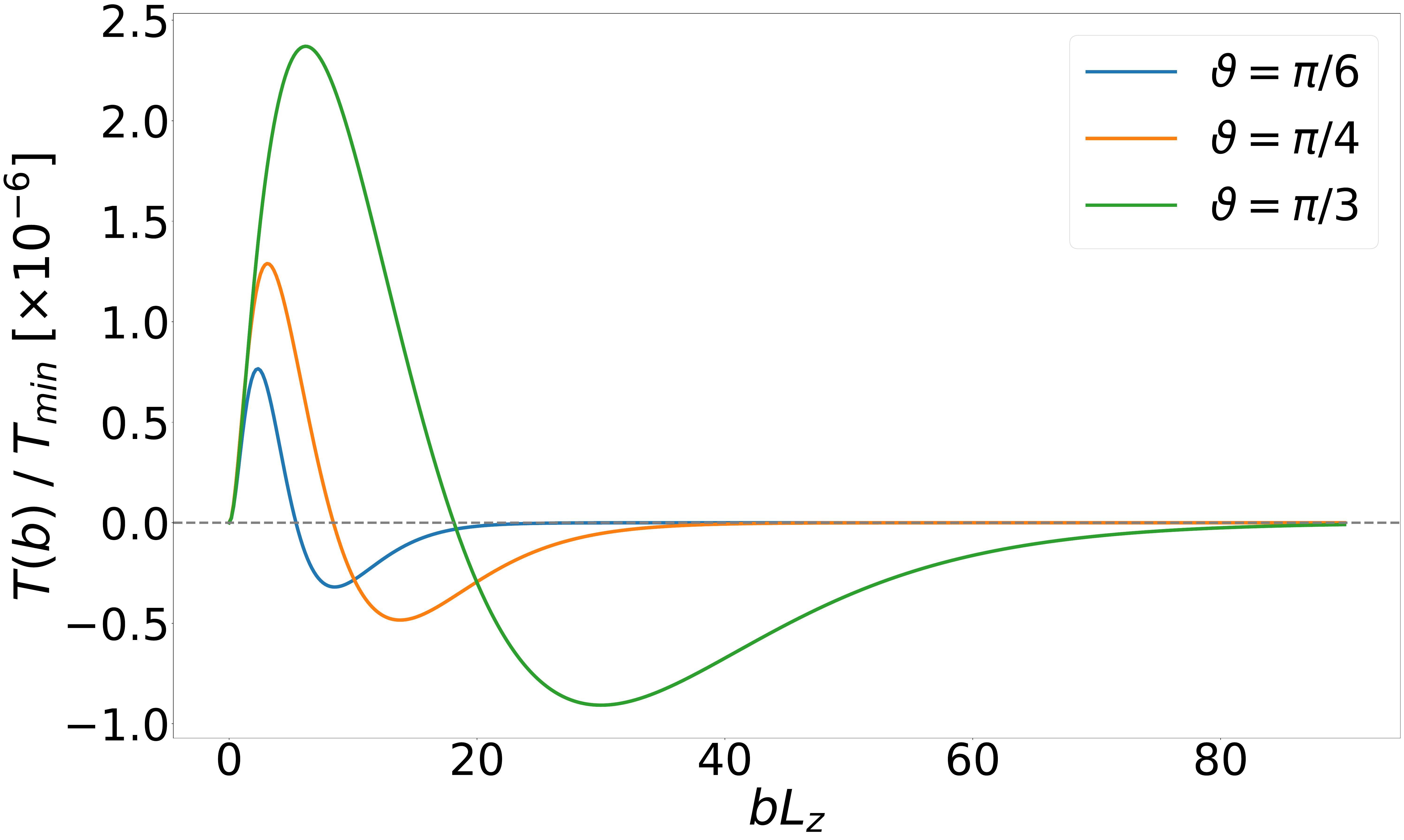}
\caption{Theoretically obtained torque normalized using parameters from Ref. \cite{Somers2018} ($T_\text{min}=3\times 10^{-9}$ N m $\text{m}^{-2} \times S$). I have used $L_z = 50 \mu \text{m}$ and $S=1 \text{mm}^2$ as sample values.}
\label{fig:T_b__vs_Tmin}
\end{figure}

To obtain the corresponding torque, we differentiate with respect to $\vartheta$ instead of $L_z$ and put on a negative sign. The general torque per unit volume is given by:
\begin{equation}
\tau = \frac{1}{\pi^2}\sum_{m=1}^{\infty}\frac{1}{Q_m^2} \int_0^{\infty}\mathrm{d}u\,u \operatorname{Re} \left[ \mathrm{e}^{-\lambda R_+} \left( A R_+ + C \right) \right] .\label{eq:20}
\end{equation}
where we have defined $\lambda = Q_m \cos \vartheta$, and $R_\pm \equiv \sqrt{u^2 \pm \mathbf{i}|\vec{b}|u}$ alongside the trigonometric auxiliary functions:
\begin{align}
A(\vartheta, u) &\equiv \frac{Q_{m} \sin(Q_{m} u \sin\vartheta)}{\cos\vartheta} , \text{ and} \label{eq:21} \\
C(\vartheta, u) &\equiv \sin(Q_{m} u \sin\vartheta) \frac{\sin^{2}\vartheta - \cos^{2}\vartheta}{\sin^{2}\vartheta \cos^{2}\vartheta} \notag \\
&+ \frac{Q_{m} u \cos(Q_{m} u \sin\vartheta)}{\sin\vartheta} .\label{eq:22}
\end{align}
We can now define $T(b) \equiv \tau (SL_z)$ and numerically analyze it (see Fig. \ref{fig:T_b__vs_Tmin}). As $u \rightarrow \infty$, the radical scales as $u$, making the exponent heavily damp the high-momentum modes. This serves as a natural UV regulator, ensuring the integral converges to a finite, observable torque density rather than blowing up into infinity. 

\begin{figure}
\centering
\includegraphics[width=1\linewidth]{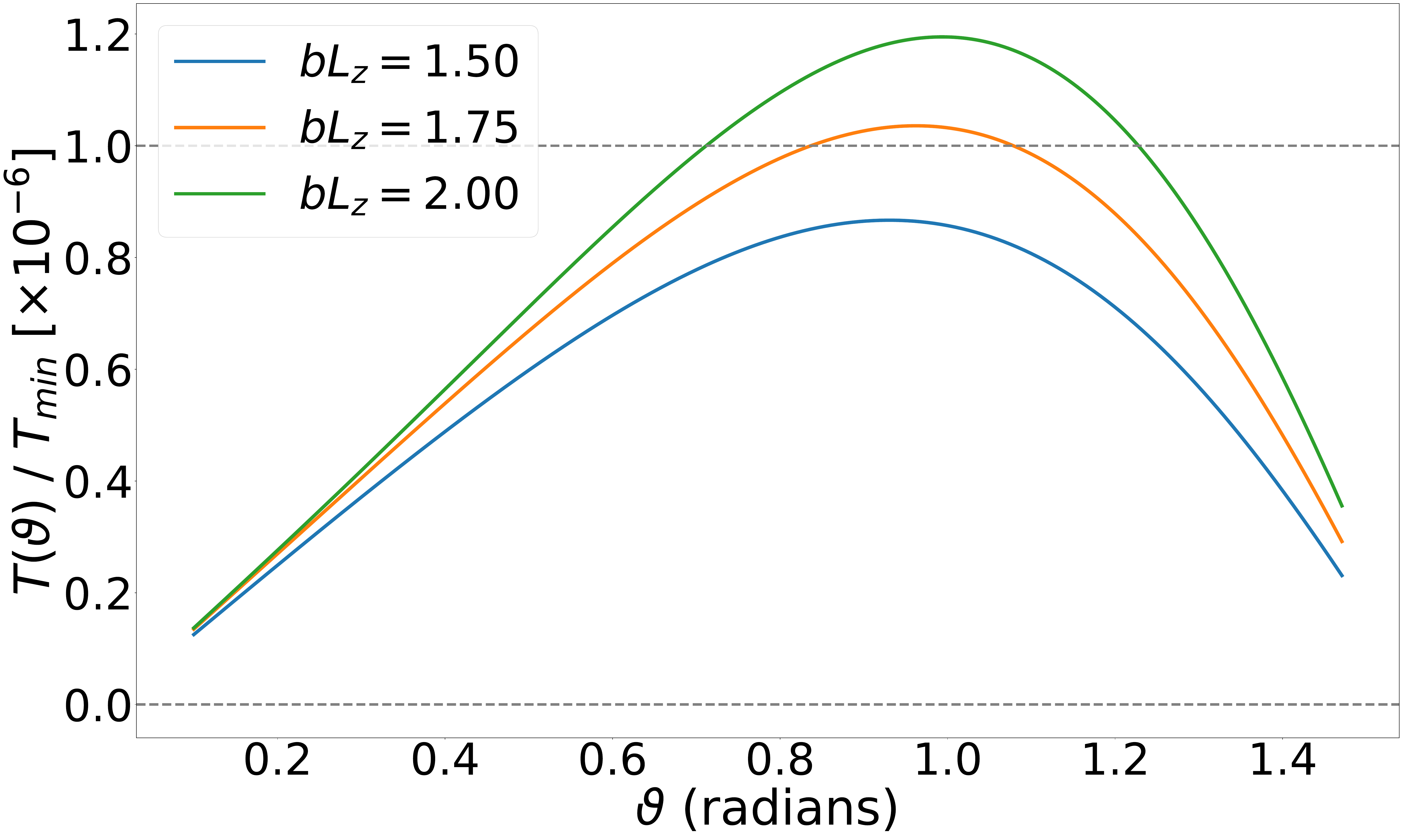}
\caption{The normalized Casimir torque $T(\vartheta)/T_\text{min}$ plotted against the misalignment angle $\vartheta$, for the same experimental parameters used for plotting Fig. \ref{fig:T_b__vs_Tmin}}
\label{fig:T(theta)vtheta}
\end{figure}

Additionally, when $\mathbf{b} \to 0$:
\begin{equation}
\lim_{\mathbf{b} \to 0} \tau = \frac{1}{\pi^2}\sum_{m=1}^{\infty}\frac{1}{Q_m^2} \int_0^{\infty}\mathrm{d}u\,u \, \mathrm{e}^{-\lambda u } (A u + C). \label{eq:23}
\end{equation}
We can formally identify this integral as a Laplace transform $L \{ \dots \}$. Evaluating the Laplace transforms, the terms analytically cancel out, yielding zero:
\begin{equation} \label{eq:24}
\lim_{\mathbf{b} \to 0} \tau = \frac{1}{\pi^2}\sum_{m=1}^{\infty} L \{ u^2 A(u) + u C(u) \} = 0
\end{equation}
This is also in agreement with the normalized numerical plot in Fig. \ref{fig:T_b__vs_Tmin}.

\subsubsection{\label{weak field limit} {Weak-field spatial torque}}

To extract the leading-order physical behavior, we consider a weak-field limit. Using the binomial theorem, we expand $R_\pm$:
\begin{equation}
R_{\pm} = u \left( 1 \pm \mathbf{i}\frac{\vec{b}}{u} \right)^{1/2} \approx u \pm \mathbf{i}\frac{\vec{b}}{2} + \frac{\vec{b}^2}{8u} .\label{eq:25}
\end{equation}
Similarly, considering the exponential decay factor:
\begin{equation}
\mathrm{e}^{-\lambda R_\pm} \approx \mathrm{e}^{-\lambda u} \left( 1 \mp \mathbf{i}\frac{\lambda \vec{b}}{2} - \frac{\lambda \vec{b}^2}{8u} - \frac{\lambda^2 \vec{b}^2}{8} \right) .\label{eq:26}
\end{equation}
Substituting Eqs. \eqref{eq:25} and \eqref{eq:26} into the polynomial bracket in Eq. \eqref{eq:20}:
\begin{equation}
A R_\pm + C \approx B \pm \mathbf{i}\frac{A \vec{b}}{2} + \frac{A \vec{b}^2}{8u}. \label{eq:27}
\end{equation}
where we define the $\mathcal{O}(b^0)$ base term as $B \equiv A u + C$. Finally:
\begin{align}
&\mathrm{e}^{-\lambda R_\pm}(A R_\pm + C) \notag \\
\approx \; &\mathrm{e}^{-\lambda u} \Big[ B + \mathbf{i}\left( \pm \frac{A}{2} \mp \frac{\lambda B}{2} \right)\vec{b} \notag \\
&\hspace{1cm} + \left( \frac{A}{8u} + \frac{\lambda A}{4} - \frac{\lambda B}{8u} - \frac{\lambda^2 B}{8} \right) \vec{b}^2  \Big] .\label{eq:28}
\end{align}
Isolating the real components yields:
\begin{align}
&\operatorname{Re}\left[ \mathrm{e}^{-\lambda R_\pm}(A R_\pm + C) \right] \notag \\
\approx \; & \underbrace{B \mathrm{e}^{-\lambda u}}_{\mathcal{O}(\vec{b}^0)} + \underbrace{\frac{\vec{b}^2}{8} \mathrm{e}^{-\lambda u} \left( - \frac{\lambda B}{u} - \lambda^2 B + 2\lambda A + \frac{A}{u} \right)}_{\mathcal{O}(\vec{b}^2)} .\label{eq:29}
\end{align}
Therefore, the $\mathcal{O}(\vec{b}^0)$ term is shown to possess no angular dependence, consistent with what we obtained so far. The $\mathcal{O}(\vec{b}^2)$ torque contribution can be integrated using Laplace transforms again to obtain:
\begin{equation} \label{eq:30}
\tau \approx \frac{\vec{b}^2 \sin(2\vartheta)}{96 L_z^2} .
\end{equation}

This regime describes condensed-matter models in which $\vec{b}$ is not a fluctuating field but is fixed by a material's magnetic or topological texture, as in axion insulators \cite{QiHughesZhang2008,EssinMooreVanderbilt2009} and Weyl semimetals \cite{WanTurnerVishwanathSavrasov2011,ArmitageMeleVishwanath2018,ZyuzinBurkov2012,ChenWuBurkov2013}, many orders of magnitude larger than in astrophysical settings. Repulsive Casimir forces have been found for topological insulators \cite{GrushinCortijo2011}, Chern insulators and quantum-Hall systems \cite{RodriguezLopezGrushin2014,TseMacDonald2012}, and Weyl semimetals \cite{WilsonAlloccaGalitski2015}, and a zero-force configuration with Weyl-semimetal-coated plates was found in Ref. \cite{Ema2023}; none of these considers a misaligned gradient. In Weyl semimetals, for instance, the axion field takes the explicitly linear form $\theta(\mathbf r,t) = 2\mathbf b\cdot\mathbf r - 2b_0 t$, where $2\mathbf b$ and $2b_0$ are fixed by the momentum- and energy-space separation of the Weyl nodes \cite{ZyuzinBurkov2012}. This could be one area to seek experimental realizations of the theory.

\section{\label{temporalaxion}A Spatio-temporal Axion}

It is of interest to examine whether a Casimir torque setup could test dark matter parameters. Modifying our assumptions to mimic astrophysical conditions, we note that cold axion dark matter has virial velocity $v\sim10^{-3}$ \cite{Evans2019,Brevik2021,Favitta2023}, so the spatial gradient is expected to be suppressed relative to $b_0$, $|\vec b|\sim b_0 v$ \cite{GrahamRajendran2011}. Therefore, we operate in the limit $b_0 \gg |\vec{b}|$ to ground our analysis in standard astrophysical phenomenology \cite{GrahamIrastorza2015}. Recall Eq. \eqref{eq:4} and the polynomial expression inside the logarithm in Eq. \eqref{eq:5}. As before, we set $\mathcal{P}(\mathbf{k})=0$ to find the roots. Expand $\mathcal{P}(\mathbf{k})$ and perform Wick's rotation to imaginary time: $k_0 \to \mathbf{i}k_4$ and $b_0 \to \mathbf{i}b_4$. This gives us:
\begin{align}
\mathcal{P}(\mathbf{k}_E) &= k_4^4 + (2\vec{k}^2 + \vec{b}^2)k_4^2 - 2b_4 k_4 (\vec{b}\cdot\vec{k}) + \vec{k}^4 \notag \\
&\hspace{1cm}+ b_4^2 \vec{k}^2 + \vec{b}^2 \vec{k}^2 - (\vec{b}\cdot\vec{k})^2. \label{eq:31}
\end{align}
Our goal is to isolate the $k_4$ poles for contour integration, so we enforce the following structure:
\begin{align}
\mathcal{P}(k_4) &= (k_4^2 - \mathbf{i} \gamma_E k_4 + \alpha_+^2)(k_4^2 + \mathbf{i} \gamma_E k_4 + \alpha_-^2)\notag \\
&= k_4^4 + (\alpha_+^2 + \alpha_-^2 + \gamma_E^2)k_4^2 \notag \\
&\hspace{2cm} - \mathbf{i} \gamma_E (\alpha_-^2  - \alpha_+^2) k_4 + \alpha_+^2 \alpha_-^2 .\label{eq:32}
\end{align}
Comparing Eqs. \eqref{eq:31} and \eqref{eq:32}, we match coefficients for $k_4^0$, $k_4^1$ and $k_4^2$. Solving the three equations, we get expressions for the unknown variables: $\alpha_\pm^2$ and $\gamma_E$. 

\begin{align} 
\gamma_E &= \pm \frac{\vec{b}\cdot\vec{k}}{\vec{k}} ,\text{ and} \label{eq:33} \\
\alpha_\pm^2 &= \vec{k}^2 \pm \mathbf{i} b_4 \vert{}\vec{k}\vert{} + \frac{\vec{b}^2}{2} - \frac{(\vec{b}\cdot\vec{k})^2}{2\vec{k}^2} .\label{eq:34}
\end{align}
Here we have dropped an $\mathcal{O}(\vec{b}^4)$ term. Using these expressions, the energy density becomes:
\begin{equation} \label{eq:35}
\mathcal{E} = \frac{1}{2L_z} \sum_\pm \sum_{n=0}^\infty \frac{\mathrm{d}k_1 \mathrm{d}k_2}{(2\pi)^2}  J_\pm ,
\end{equation}
where:
\begin{align}
J_\pm &\equiv \int_{-\infty}^{\infty} \frac{\mathrm{d}k_4}{2\pi} \ln(k_4^2 \mp \mathbf{i} \gamma_E k_4 + \alpha_\pm^2) \notag \\
&= \int_{-\infty}^{\infty} \frac{\mathrm{d}k_4}{2\pi} \ln \left[ \left( k_4 \mp \mathbf{i} \frac{\gamma_E}{2} \right)^2 + \Omega_\pm^2 \right],  \label{eq:36}
\end{align}
and:
\begin{equation} \label{eq:37}
\Omega^2_\pm \equiv \alpha_\pm^2 + \frac{\gamma_E^2}{4} = \vec{k}^2 \pm b_0 \vert{}\vec{k}\vert{} + \frac{\vec{b}^2}{2} - \frac{(\vec{b}\cdot\vec{k})^2}{4\vec{k}^2}.
\end{equation}
A timelike $b_\mu$ is known to allow tachyonic long-wavelength modes \cite{Fukushima2019}; here $\Omega_-^2<0$ for $|\vec k|<b_0$. Ref. \cite{KharlanovZhukovsky2010} found that this instability is absent for sufficiently small plate separation. For $b_0L_z\ll\pi$, every mode with $k_3=n\pi/L_z$, $n\ge1$, has $|\vec k|>b_0$, so $\Omega_\pm^2>0$.

\subsection{\label{energy2}Spatio-temporal Energy}

To evaluate this, we shift the integration contour to $u \equiv k_4 \mp \mathbf{i} \frac{\gamma_E}{2}$. Cauchy's integral theorem guarantees the invariance of this shift. The positive spatial momentum terms and the Cauchy-Schwarz bound on the axion gradient prevent the argument of the logarithm from crossing the principal branch cut $(-\infty, 0]$ (see Appendix \ref{AppendixA} for the proof). Therefore, Eq. \eqref{eq:36} becomes:
\begin{equation} \label{eq:38}
J_\pm (\Omega_\pm) = \int_{-\infty}^{\infty} \frac{\mathrm{d} u}{2\pi} \ln \left[ u^2 + \Omega_\pm^2 \right] = \Omega_\pm + C_\infty.
\end{equation}
The integration constant $C_{\infty}$ is discarded (renormalized to zero). Substituting this into Eq. \eqref{eq:35}, we obtain:
\begin{equation} \label{eq:39}
\mathcal{E} = \frac{1}{2L_z} \sum_\pm \sum_{n=0}^\infty \int \frac{\mathrm{d}k_1 \mathrm{d}k_2}{(2\pi)^2} \left( \vec{k}^2 \pm \mathbf{i} b_4 \vert{}\vec{k}\vert{} + \frac{\vec{b}^2}{2} - \frac{(\vec{b}\cdot\vec{k})^2}{4\vec{k}^2} \right)^{1/2} .
\end{equation}
Analytically continuing the temporal background field back to Minkowski space using $b_0 = \mathbf{i}b_4$ mathematically mandates $b_4 = -\mathbf{i}b_0$. 
\begin{equation} \label{eq:40}
\mathcal{E} = \frac{1}{2L_z}  \sum_{n=0}^\infty \int \frac{\mathrm{d}k_1 \mathrm{d}k_2}{(2\pi)^2} \sum_\pm \Omega_\pm .
\end{equation}
Bundle the transverse integral into a function $F(n)$ dependent entirely on the mode integer $n$:
\begin{equation} \label{eq:41}
F(n) = \int \frac{\mathrm{d}k_1 \mathrm{d}k_2}{(2\pi)^2} \sum_{\pm} \Omega_\pm .
\end{equation}
Inverting the longitudinal mode $n \rightarrow -n$ maps $k_3 \rightarrow -k_3$, which alters the symmetry-breaking scalar product $(\vec{b}\cdot\vec{k})^2$. However, this is offset by a parity inversion of the transverse integration variables, $(k_1, k_2) \rightarrow (-k_1, -k_2)$. The infinite integration measure is invariant, and the scalar product is recovered: $(-b_1 k_1 - b_2 k_2 - b_3 k_3)^2 = (\vec{b}\cdot\vec{k})^2$. This confirms $F(n) = F(-n)$. Because it is even, we may now apply the Poisson summation formula:
\begin{equation} \label{eq:42}
\sum_{n=0}^\infty F(n) = \frac{1}{2} \sum_{n=-\infty}^{\infty} F(n) = \frac{1}{2} \sum_{m=-\infty}^{\infty} \int_{-\infty}^{\infty} \mathrm{d}n \, F(n) e^{2\pi \mathbf{i} m n} .
\end{equation}
Substituting this expansion back into the total energy equation yields:
\begin{equation} \label{eq:43}
\mathcal{E} = \frac{1}{4L_z} \sum_{m=-\infty}^{\infty} \int_{-\infty}^{\infty} \mathrm{d}n \int \frac{\mathrm{d}k_1 \mathrm{d}k_2}{(2\pi)^2} \sum_{\pm} \Omega_\pm e^{2\pi \mathbf{i} m n}.
\end{equation}
We now transition from the mathematical mode number $n$ to the physical longitudinal momentum $k_3 = \frac{n\pi}{L_z}$.
\begin{equation} \label{eq:44}
\mathcal{E} = \frac{1}{2} \sum_{m=-\infty}^{\infty} \int \frac{\mathrm{d}^3 k}{(2\pi)^3} \sum_{\pm} \Omega_\pm \, e^{\mathbf{i} Q_m k_3},
\end{equation}
where, $Q_m = 2mL_z$. The specific harmonic corresponding to $m=0$ maps to the unbounded continuum bulk vacuum energy density. Because it lacks the exponential phase, it is independent of the physical plate separation $L_z$. Furthermore, using $\mathrm{e}^{\mathbf{i} Q_m k_3} = \cos(Q_m k_3) + \mathbf{i} \sin(Q_m k_3)$, the imaginary sine component is odd and thus integrates to zero.
\begin{equation} \label{eq:45}
\mathcal{E}_{\text{Casimir}} = \sum_{m =1}^\infty \int \frac{\mathrm{d}^3 k}{(2\pi)^3} \left(\Omega_+ + \Omega_-\right) \cos(Q_m k_3).
\end{equation}

\subsubsection{Recovering the canonical result}

Taking the limits $b_0 \to 0$ and $\vec{b} \to 0$, we find $\Omega_\pm(b \to 0) = | \vec{k} |$. The sum over the two polarization branches yields $ 2\vert{}\vec{k}\vert{}$. Converting to spherical coordinates gives:
\begin{align}
\mathcal{E} (0) &= \sum_{m=1}^{\infty} \int \frac{\mathrm{d}^3 k}{(2\pi)^3} (2\vert{}\vec{k}\vert{}) \cos(Q_m k_3) \notag \\
&= \frac{1}{2\pi^2L_z}\sum_{m=1}^{\infty} \frac{1}{m} \int_0^\infty \mathrm{d}\kappa \, \kappa^2 \sin(Q_m \kappa).\label{eq:46}
\end{align}
This radial momentum integral oscillates over an infinite domain and is formally divergent. Introduce a factor $e^{-\epsilon \kappa}$.
\begin{equation} \label{eq:47}
\mathcal{E}(0) = \lim_{\epsilon \to 0} \frac{1}{2\pi^2L_z}\sum_{m=1}^{\infty} \frac{1}{m} \int_0^\infty \mathrm{d}\kappa \, \kappa^2 \sin(Q_m \kappa) e^{-\epsilon \kappa}.
\end{equation}
This is equivalent to evaluating the second derivative of the fundamental Laplace transform with respect to the damping parameter $\epsilon$. Under the limit $\epsilon \to 0$, this evaluates to the canonical result: $\mathcal{E}(0) = - \frac{\pi^2}{720L_z^4}$.

\subsection{\label{force2}Spatio-temporal Force}

The Casimir force per unit volume can be written as $\mathcal{F}(b) = - \partial \mathcal{E}_\text{Casimir}/\partial L_z$
\begin{equation} \label{eq:48}
\mathcal{F} (\mathbf{b})=\frac{1}{4\pi^3} \sum_{m=1}^{\infty} m \int \mathrm{d}^3k \; k_3 \sum_{\pm} \Omega_\pm \sin(Q_m k_3) .
\end{equation}
When $\mathbf{b} \to 0$, on converting to spherical coordinates and integrating:
\begin{equation}
\mathcal{F} (0) = \frac{1}{2\pi^3} \sum_{m=1}^{\infty} m \int \mathrm{d}^3k \; k_3 |\vec{k}| \sin(Q_m k_3) \notag =  -\frac{\pi^2}{180 L_z^5}.\label{eq:49}
\end{equation}
Let $E$ be the total energy, $A$ the transverse area of the plates, and $L_z$ the separation distance. The physical pressure on the plates is:
\begin{align}
\frac{F}{A} &= -\frac{\partial}{\partial L_z} (\mathcal{E} L_z) = -\mathcal{E} - L_z \frac{\partial \mathcal{E}}{\partial L_z} \notag \\
&= -\left( - \frac{\pi^2}{720 L_z^4} \right) - L_z \left( \frac{\pi^2}{180 L_z^5} \right) =  - \frac{\pi^2}{240 L_z^4}.\label{eq:50}
\end{align}
Therefore, the theory remains consistent.

\subsection{\label{torque2}Spatio-temporal Torque}

The torque can be obtained by: $\tau (\mathbf{b}) = - \partial \mathcal{E}_\text{Casimir}/\partial \vartheta$. Due to the continuous azimuthal symmetry in the transverse $(k_1, k_2)$ plane, we orient the spatial axion gradient entirely within the $k_1$-$k_3$ plane without loss of generality, setting $\vec{b} = \vert{}\vec{b}\vert{}(\sin\vartheta, 0, \cos\vartheta)$. Taking the angular derivative of the eigenvalues $\Omega_\pm$ yields:
\begin{equation} \label{eq:51}
\frac{\partial \Omega_\pm}{\partial \vartheta} = - \frac{\vec{b}^2 \left[ (k_1^2 - k_3^2)\sin(2\vartheta) + 2k_1 k_3 \cos(2\vartheta) \right]}{8\vec{k}^2 \Omega_\pm}.
\end{equation}
The partial derivative operator passes cleanly inside the integral:
\begin{eqnarray} 
\tau &=& \frac{\vec{b}^2}{8} \sum_{m=1}^{\infty} \int \frac{\mathrm{d}^3 k}{(2\pi)^3} \sum_{\pm} \frac{(k_1^2 - k_3^2)\sin(2\vartheta) + 2k_1 k_3 \cos(2\vartheta)}{\vec{k}^2 \sqrt{ \vec{k}^2 \pm b_0 \vert{}\vec{k}\vert{} + \frac{\vec{b}^2}{2} - \frac{(\vec{b}\cdot\vec{k})^2}{4\vec{k}^2} }} \notag \\
& & \times \cos(Q_m k_3) .\label{eq:52}
\end{eqnarray}

\subsubsection{\label{b0ggb}Heavy temporal limit}

To further simplify Eq. \eqref{eq:52}, we analyze the denominator.
We begin by defining $\Omega^2_\pm = \Omega^{(0)}_\pm + \Omega^{(2)}_\pm$ where:
\begin{align}
\Omega^{(0)}_\pm &=  \vec{k}^2 \pm b_0 \vert{}\vec{k}\vert{} ,\text{ and}\label{eq:53}\\
\Omega^{(2)}_\pm &= \frac{\vec{b}^2}{2} - \frac{(\vec{b}\cdot\vec{k})^2}{4\vec{k}^2} .\label{eq:54}
\end{align}
We perform a formal binomial expansion of the inverse denominator:
\begin{equation} \label{eq:55}
\frac{1}{\vec{k}^2 \Omega_\pm} = \frac{1}{\vec{k}^2 \sqrt{\Omega^{(0)}_\pm}} - \frac{\Omega^{(2)}_\pm}{2\vec{k}^2 \left( \sqrt{\Omega^{(0)}_\pm}\right)^{3}} + \mathcal{O}(\vec{b}^4).
\end{equation}
Substituting this expansion back into the Casimir torque expression generates two primary integrals, $\tau = I_1 + I_2 + \mathcal{O}(\vec{b}^4)$ where:
\begin{widetext}
\begin{align}
I_1 &= \frac{\vec{b}^2}{8} \sum_{m=1}^{\infty} \int \frac{\mathrm{d}^3 k}{(2\pi)^3} \sum_{\pm} \frac{(k_1^2 - k_3^2)\sin(2\vartheta) + 2k_1 k_3 \cos(2\vartheta)}{\vec{k}^2 \sqrt{\Omega^{(0)}_\pm}}  \cos(Q_m k_3), \text{ and} \label{eq:56}\\
I_2 &= -\frac{\vec{b}^2}{8} \sum_{m=1}^{\infty} \int \frac{\mathrm{d}^3 k}{(2\pi)^3} \sum_{\pm} \frac{(k_1^2 - k_3^2)\sin(2\vartheta) + 2k_1 k_3 \cos(2\vartheta)}{2\vec{k}^2 \left( \sqrt{\Omega^{(0)}_\pm}\right)^{3}} \; \Omega^{(2)}_\pm \cos(Q_m k_3). \label{eq:57}
\end{align}
\end{widetext}
The integration domain $\mathbb{R}^3$ is completely symmetric. In $I_1$, the cross-term $2k_1 k_3 \cos(2\vartheta)$ is odd under $k_1 \rightarrow -k_1$ and thus vanishes entirely upon integration. For the correction term $I_2$, the perturbation $\Omega^{(2)}_\pm = \frac{\vec{b}^2}{2} - \frac{\vec{b}^2}{4\vec{k}^2} (k_1^2 \sin^2\vartheta + 2k_1 k_3 \sin\vartheta \cos\vartheta + k_3^2 \cos^2\vartheta)$ contains an odd component proportional to $k_1 k_3$. When multiplied by the odd cross-term in the numerator of $I_2$, it yields an even function that does not integrate to zero. However, because $\Omega^{(2)}_\pm$ carries an explicit factor of $\vec{b}^2$ and $I_2$ possesses an overall prefactor of $\frac{\vec{b}^2}{8}$, these non-vanishing parity-breaking terms contribute exclusively at $\mathcal{O}(\vec{b}^4)$ and higher.  Applying the \textit{strong temporal background limit} $b_0 \gg \vert\vec{b}\vert$, we truncate to leading order in $\vec{b}^2$, discarding $I_2$ and all higher-order corrections.
\begin{equation} \label{eq:58}
\tau \approx \frac{\vec{b}^2}{8} \sin(2\vartheta) \sum_{m=1}^{\infty} \int \frac{\mathrm{d}^3 k}{(2\pi)^3} \sum_{\pm} \frac{k_1^2 - k_3^2}{\vec{k}^2 \sqrt{ \vec{k}^2 \pm b_0 \vert{}\vec{k}\vert{} }} \cos(Q_m k_3).
\end{equation}
Transforming to spherical momentum coordinates $(\kappa, \theta_k, \phi_k)$ and integrating over the azimuthal angle $\phi_k \in [0, 2\pi]$ yields a factor of $\pi \kappa^2 (1 - 3\cos^2\theta_k)$ from the numerator. The remaining polar integral over $\theta_k$ evaluates directly to the spherical Bessel function $j_2(Q_m \kappa)$, allowing the torque to be compactly written as:
\begin{equation} \label{eq:59}
\tau \approx \frac{\vec{b}^2}{16\pi^2} \sin(2\vartheta) \sum_{m=1}^{\infty} \int_0^\infty \mathrm{d}\kappa \sum_{\pm} \frac{\kappa^2}{\sqrt{ \kappa^2 \pm b_0 \kappa }}j_2 (Q_m \kappa).
\end{equation}
To further simplify, we can isolate the term inside the integral containing the temporal gradient and perform a binomial expansion for $\kappa > b_0$. The Casimir torque is a macroscopic manifestation of vacuum mode quantization, where the dominant momentum contributions are set by the cavity scale, $\kappa \sim \pi/L_z \implies b_0 L_z \ll \pi$. To extract the leading-order physical behavior, we treat the integrand of Eq. \eqref{eq:59} as a formal asymptotic expansion in the small parameter $b_0/\kappa$. Extending this expansion over the entire integration domain $[0, \infty)$ formally introduces artificial infrared singularities in higher-order terms at $\kappa \to 0$. However, because the physical vacuum modes are suppressed by the spherical Bessel function $j_2(Q_m \kappa)$, these divergences are mathematical artifacts of the expansion rather than physical instabilities.

Defining an auxiliary function and applying the binomial expansion yields: 
\begin{align}
f(k) &\equiv \sum_{\pm} \frac{\kappa^2}{\sqrt{\kappa^2 \pm b_0 \kappa}} \notag \\
&= \kappa \sum_{p=0}^\infty \left[ \binom{-1/2}{p}\left( \frac{b_0}{\kappa}\right)^p +\binom{-1/2}{p}\left( -\frac{b_0}{\kappa}\right)^p \right] \notag \\
&=  \kappa \sum_{p=0}^\infty \underbrace{\left( \frac{2 \Gamma\left(2p + \frac{1}{2}\right)}{\sqrt{\pi} \Gamma (2p+1)} \right)}_{c_p} \left(\frac{b_0}{\kappa}\right)^{2p} = \sum_{p=0}^\infty c_p \kappa^{1-2p} b_0^{2p} .\label{eq:60}
\end{align}
The first three coefficients evaluate to: $c_0 = 2$, $c_1 = \frac{3}{4}$, and $c_2 = \frac{35}{64}$. Substituting this expansion back into Eq. \eqref{eq:59} yields:
\begin{equation}
\tau \approx \frac{\vec{b}^2}{16\pi^2} \sin(2\vartheta)  \sum_{p=0}^\infty c_p b_0^{2p} \sum_{m=1}^{\infty} \int_0^\infty \mathrm{d}\kappa \; \kappa^{1-2p}  j_2 (Q_m \kappa) .\label{eq:61}
\end{equation}
Define $s \equiv 2-2p$ such that $s-1 = 1-2p$. Let us evaluate the general function:
\begin{align}
L(s) &= \sum_{m=1}^{\infty} \int_0^\infty \mathrm{d}\kappa \, \kappa^{s-1} j_2(Q_m \kappa) \label{eq:62} \\
&= \frac{\sqrt{\pi}}{4 L_z^s} \frac{\Gamma\left(1 + \frac{s}{2}\right)}{\Gamma\left(\frac{5}{2} - \frac{s}{2}\right)} \zeta(s) .\label{eq:63}
\end{align}
Revisiting the integral in Eq. \eqref{eq:62}, $\int_0^\infty \mathrm{d}\kappa \, \kappa^{s-1} j_2(Q_m \kappa)$, we substitute $x\equiv Q_m \kappa$ to obtain:
\begin{align}
\frac{1}{Q_m^s} \int_0^{\infty} x^{s-1} j_2(x) \mathrm{d}x .\label{eq:64}
\end{align}
Reincorporating the infinite summation over the longitudinal modes $m$:
\begin{align}
L(s) &= \sum_{m=1}^{\infty} \frac{1}{(2mL_z)^s} \int_0^{\infty} x^{s-1} j_2(x) \mathrm{d}x \notag \\
&= \frac{\zeta(s)}{2^s L_z^s} \int_0^{\infty} x^{s-1} j_2(x) \mathrm{d}x .\label{eq:65}
\end{align}

Comparing Eqs. \eqref{eq:63} and \eqref{eq:65}:
\begin{equation} \label{eq:66}
\int_0^{\infty} x^{s-1} j_2(x) \mathrm{d}x = 2^{s-2}\sqrt{\pi} \frac{\Gamma(1+\frac{s}{2})}{\Gamma(\frac{5}{2}-\frac{s}{2})}.
\end{equation}

\begin{figure}[htbp]
\centering 
\includegraphics[width=1\linewidth]{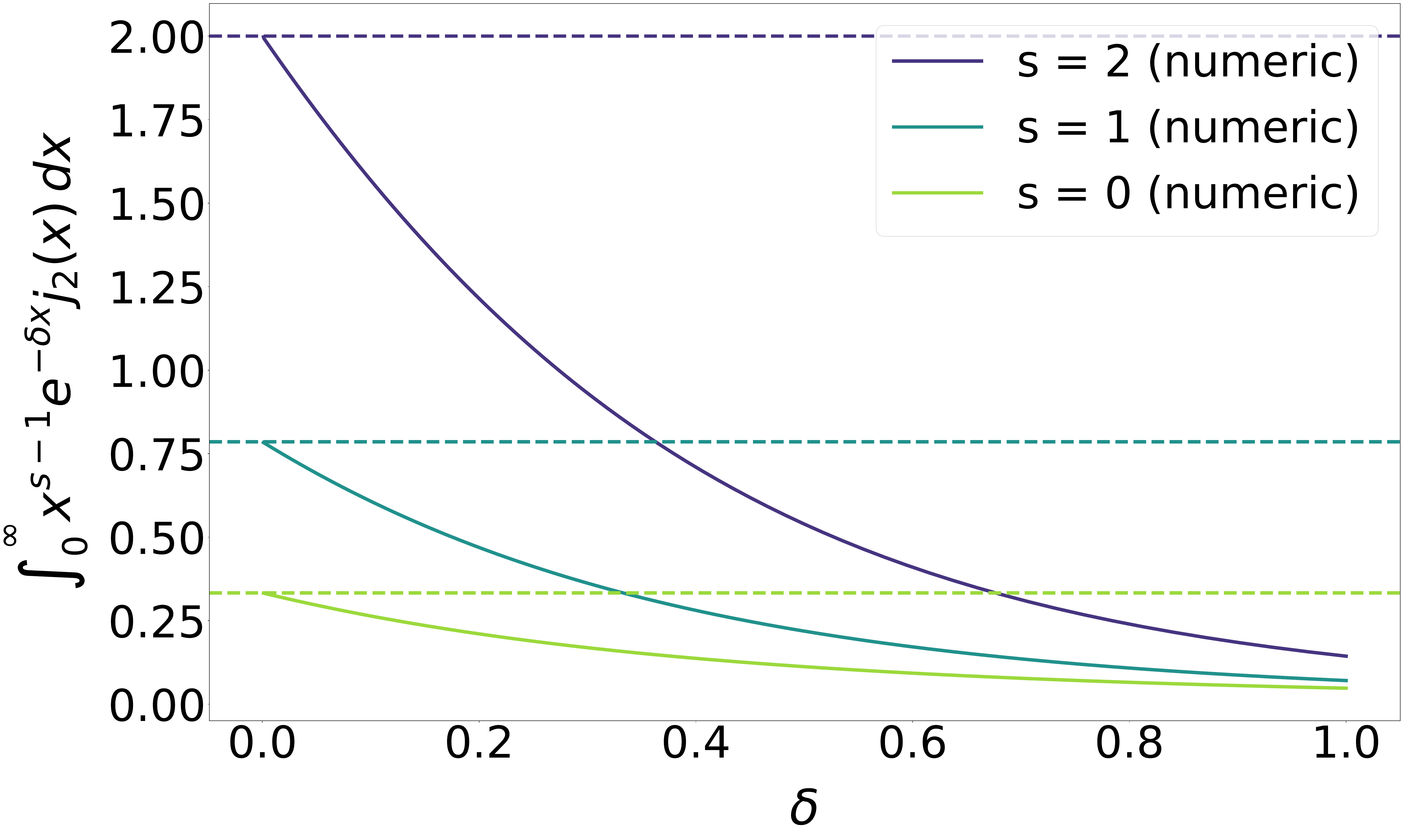}
\caption{The dashed lines indicate the exact closed-form analytical limits derived via the continuous Gamma function for $s=2, 1$, and $0$. The solid curves trace the direct numerical integration of the spherical Bessel function $j_2(x)$, stabilized by an exponential  damping parameter $e^{-\delta x}$. As the unphysical damping parameter vanishes ($\delta \to 0$), the regulated numerical solutions smoothly converge to the analytical limits.}
\label{fig:MellinCheck}
\end{figure}

We finally evaluate $L(s)$ explicitly for the first three terms of our $b_0$ expansion where $s = 2-2p$.

\begin{figure}[htbp]
\centering
\includegraphics[width=1\linewidth]{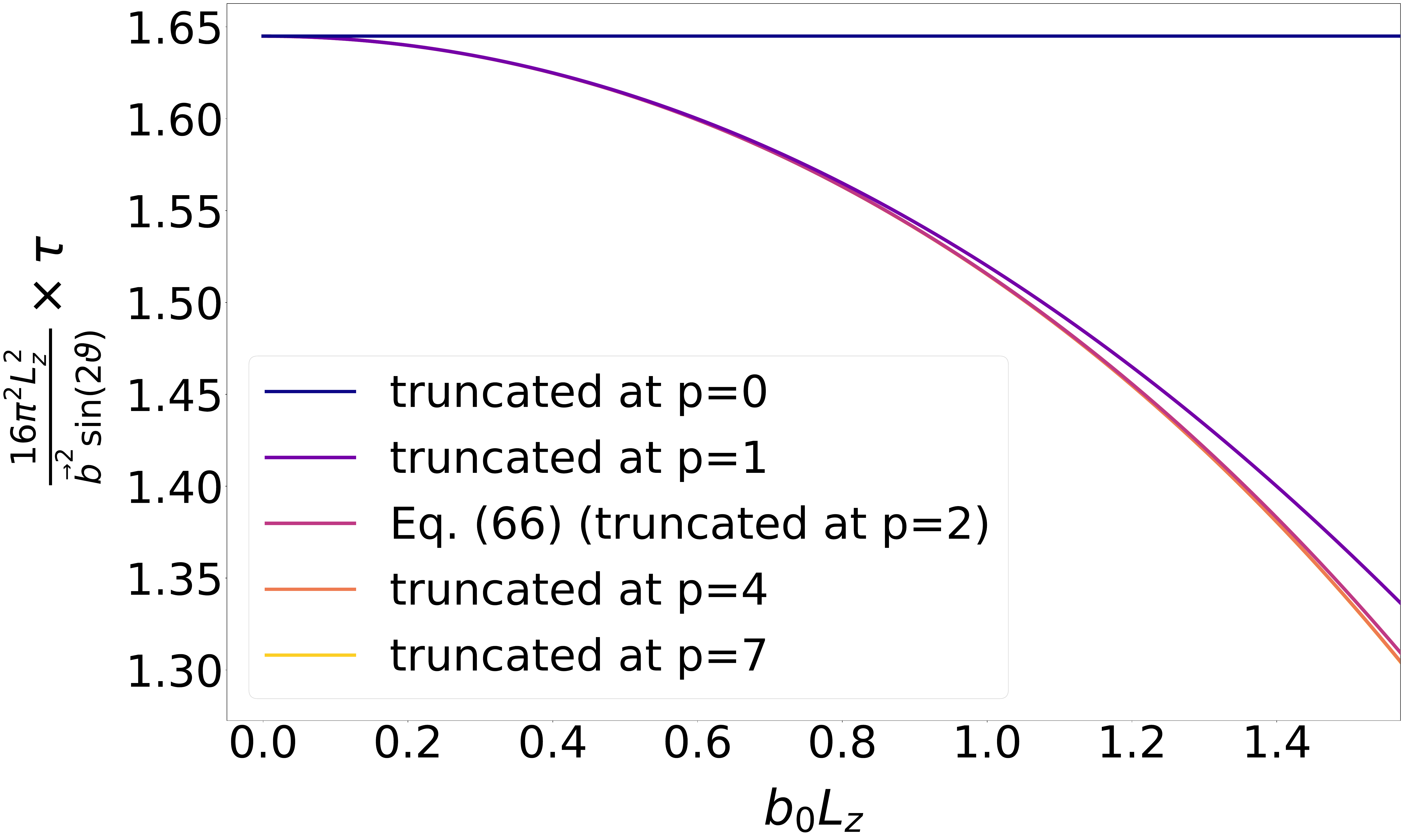}
\caption{The normalized spatio-temporal Casimir torque coefficient, $\tau \times \frac{16\pi^2 L_z^2}{\vec{b}^2 \sin(2\vartheta)}$, plotted against $b_0 L_z$. The curves demonstrate the behavior of the asymptotic expansion derived in Eq. \eqref{eq:67}, truncated at successive orders $p \in \{0, 1, 2, 4, 7\}$. When $b_0 L_z \lesssim 0.5 \ll \pi$, the series is well-behaved and successive orders provide rapidly converging corrections to the spatial base term.}
\label{fig:pcheck}
\end{figure}

\begin{itemize}
\item[i)] \textbf{leading order} ($p=0 \implies s=2$): $L(2) = \frac{\pi^2}{12 L_z^2}$, yielding the contribution: $c_0 b_0^0 L(2) = \frac{\pi^2}{6 L_z^2}$.
\item[ii)] \textbf{first temporal correction} ($p=1 \implies s=0$): $L(0) = -\frac{1}{6}$, yielding the contribution $c_1 b_0^2 L(0) = -\frac{b_0^2}{8}$.  
\item[iii)] \textbf{second temporal correction} ($p=2 \implies s=-2$): The limit as $s \to -2$ requires analytic continuation. By taking the product of the Taylor series expansion $\zeta(s) \approx -\frac{\zeta(3)}{4\pi^2}(s+2)$ and the Laurent series of the Gamma function near its pole, the limit evaluates to $L(-2) = -\frac{L_z^2 \zeta(3)}{15\pi^2}$. The contribution is therefore $c_2 b_0^4 L(-2) = -\frac{7 L_z^2 b_0^4 \zeta(3)}{192 \pi^2}$.  
\end{itemize}
Summing the regularized contributions yields:
\begin{equation}\label{eq:67}
\tau \approx \sin(2\vartheta) \left[ \frac{\vec{b}^2}{96 L_z^2} - \frac{\vec{b}^2 b_0^2}{128 \pi^2} - \frac{7 \zeta(3) \vec{b}^2 b_0^4 L_z^2}{3072 \pi^4} \right] .
\end{equation}
Setting $b_0 = 0$ recovers Eq. \eqref{eq:30}. A longer series can be considered by increasing the value of $p$ (see Fig. \ref{fig:pcheck}), but higher-order terms rapidly converge, yielding little noticeable difference.

\section{\label{conclusions}Conclusions}

We have thus shown that a misaligned axion background gives rise to a torque with the same $\sin(2\vartheta)$ functional form seen for anisotropic dielectrics \cite{Somers2018,XuLi2017}. We can reproduce the leading spatial and spatio-temporal terms using a Schwinger proper-time evaluation (Appendix \ref{AppendixB}). Both results reduce to the canonical, angle-independent Casimir energy $-\pi^{2}/720L_{z}^{4}$ as $\mathbf{b}\to0$, and the torque itself vanishes in that limit, consistent with the absence of a preferred direction once the axion gradient is
removed. All plots are generated using Python and the codes are available on request.

A larger body of work has proposed, but not yet carried out, vacuum-gap torque measurements accompanied by quantitative sensitivity estimates. Early torsion-balance designs estimated a birefringent-plate torque of order $10^{-15}\,\mathrm{N\,m}$ for two $\mathrm{LiNbO}_3$ disks of 1 cm radius at 1 $\mu$m separation, benchmarked against a stated apparatus sensitivity of order $10^{-17}\,\mathrm{N\,m}$ \cite{Iannuzzi2003}.  Ref. \cite{Guerout2015} subsequently found that at 100 nm separation the torque between gold gratings can be about three orders of magnitude larger than for birefringent dielectric plates at the same distance, and proposed a torsion-balance scheme to measure it. Ref. \cite{XuLi2017} proposed detecting the torque via an optically levitated nanorod near a birefringent plate. They projected a torque-detection sensitivity of order $10^{-28}\,\mathrm{N\,m/\sqrt{Hz}}$ for a 200 nm $\times$ 40 nm silica nanorod at $10^{-7}$ Torr, against a calculated torque of order $10^{-25}\,\mathrm{N\,m}$ at 266 nm separation. They benchmarked this against the best torque sensitivity reported at the time, $2.9\times10^{-24}\,\mathrm{N\,m/\sqrt{Hz}}$, obtained from a cavity-optomechanical torque sensor operated at 25 mK.  Ref. \cite{Ahn2018} reported the synthesis and optical levitation of silica nanodumbbells in high vacuum and observed their torsional vibration; they calculated a torque sensitivity of $\sim10^{-27}\text{ N}\cdot\text{m}/\sqrt{\text{Hz}}$ ($D=170$ nm) and $\sim10^{-29}\text{ N}\cdot\text{m}/\sqrt{\text{Hz}}$ ($D=50$ nm) below $10^{-8}$ torr. Ref. \cite{Ahn2019} further utilized a fast-rotating optically levitated silica nanoparticle in vacuum. They observed rotation beyond 5 GHz, demonstrated a room-temperature torque sensitivity of a few $10^{-27}\text{ N}\cdot\text{m}/\sqrt{\text{Hz}}$, and calculated that the vacuum-friction torque on a 75-nm-radius silica sphere rotating at 1 GHz near a silica surface at 1000 K approaches $10^{-27}\text{ N}\cdot\text{m}$ at small separations. Additionally, Ref. \cite{Hu2025} computed the spectral Casimir friction torque and radiative heat flux for a rotating Weyl-semimetal nanoparticle above a plate ($M_\omega\sim10^{-44}$--$10^{-50}\text{ N}\cdot\text{m}\cdot\text{s/rad}$); this is a calculated spectral torque in the context of energy conversion, not a detection sensitivity. A comprehensive review, covering both the proposed and realized configurations, is given by Ref. \cite{Spreng2022}. 

None of these experimental setups, however, utilize the exact geometry described in this work. To date, direct observations of Casimir-induced torques have bypassed traditional vacuum gaps to mitigate parallelism challenges. Ref. \cite{Somers2018} measured the torque between a solid birefringent crystal and a liquid-crystal layer by separating a solid birefringent crystal from a nematic liquid crystal (5CB) with an optically isotropic Al$_2$O$_3$ spacer (0--25 nm) plus a $\sim$12 nm surfactant layer, i.e.\ at effective separations of tens of nanometres. In a related physical demonstration, Ref.  \cite{Kucukoz2024} observed the Casimir-driven rotational self-alignment of triangular gold nanostructures at 100--200 nm separations. In this setup, a stable electrostatic-Casimir trap self-aligns to maximize the overlap area between two flakes. This serves as a demonstration of the torque acting as a self-assembly driver rather than a direct torque readout, a result that a subsequent comment noted has overlap with earlier theoretical treatments of tangential Casimir forces in misaligned systems \cite{zhang2024Comment}. Using the Casimir effect as a probe of the background axion angle has been proposed for a static $\theta$ in a system with non-trivial spatial topology \cite{CaoZhitnitsky2017}. 

We evaluate the theoretical torque using standard dark matter halo parameters, where the spatial gradient is suppressed by $\sim 10^{-3}$. For a macroscopic cavity with a $1 \text{ mm}^2$ area and a $50 \text{ }\mu\text{m}$ separation, this yields a predicted mechanical torque of order $\approx 10^{-65}\text{ N m}$ (or $\sim 10^{-56}\text{ GeV}$), beyond current sensitivities.

\begin{acknowledgments}
I would like to thank Subhendra Mohanty for his suggestions, encouragement, and ideas during the preparation of this manuscript.
\end{acknowledgments}

\appendix

\section{\label{AppendixA}Validity of the contour shift}

Any $z \in \mathbb{C}$ can be written as $z = r e^{\mathbf{i}(\theta + 2\pi n)}$ $\forall$ $n\in \mathbb{Z}$. Consequently, the complex logarithm has infinitely many valid outputs for a single input: $\ln(z) = \ln(r) + \mathbf{i}(\theta + 2\pi n)$. To force the multi-valued complex logarithm to behave as a standard function, we must mathematically forbid it from completing a full circle around the origin. We enforce this by drawing a ``barrier''---the branch cut---across which we cannot integrate. 

We have $Z_\pm(k_4) = (k_4 \mp \mathbf{i} \frac{\gamma_E}{2})^2 + \Omega_\pm^2$. To ensure the contour shift $u \equiv k_4 \mp \mathbf{i} \frac{\gamma_E}{2}$ avoids the principal branch cut $(-\infty, 0]$, we prove that $\mathrm{Re}(Z_\pm) > 0$ globally. Parameterizing $k_4 = x + \mathbf{i}y$ and defining the bounded translation parameter $v \equiv y \mp \frac{\gamma_E}{2} \in [\mp \frac{\gamma_E}{2}, 0]$, we substitute $k_4 = x + \mathbf{i}\left(v \pm \frac{\gamma_E}{2}\right)$ into $Z_\pm$:
\begin{equation}
Z_\pm(x, v) = \left( x^2 - v^2 + \vec{k}^2 + \frac{1}{2}\vec{b}^2 - \frac{\gamma_E^2}{4} \right) + \mathbf{i} \left( 2xv \pm b_4 \vert\vec{k}\vert \right) . \label{eq:A1}
\end{equation}
Extracting the real part and applying the bounds of the translation contour, $v^2 \leq \frac{\gamma_E^2}{4}$, establishes a lower bound:
\begin{equation}
\mathrm{Re}(Z_\pm) \geq x^2 + \vec{k}^2 + \frac{1}{2}\vec{b}^2 - \frac{\gamma_E^2}{2}. \label{eq:A2}
\end{equation}
From Eq. \eqref{eq:33}, using the Cauchy-Schwarz inequality, $\gamma_E^2 = \frac{(\vec{b}\cdot\vec{k})^2}{\vec{k}^2} \leq \frac{\vec{b}^2 \vec{k}^2}{\vec{k}^2} = \vec{b}^2$. Multiplying by $-1/2$ and substituting this bound into Eq. \eqref{eq:A2} yields:
\begin{equation}
\mathrm{Re}(Z_\pm) \geq x^2 + \vec{k}^2 + \frac{1}{2}\vec{b}^2 - \frac{\vec{b}^2}{2}= x^2 + \vec{k}^2. \label{eq:A3}
\end{equation}
Because $x \in \mathbb{R}$ implies $x^2 \geq 0$, and the physical phase space measure excludes the trivial $\vec{k}=0$ state, $\vec{k}^2 > 0$. We therefore conclude: $\mathrm{Re}(Z_\pm) > 0$. Since $\mathrm{Re}(Z_\pm)$ is positive across the entire parameter space of the shift, $Z_\pm$ is confined to the right half-plane. The branch cut $(-\infty, 0]$ is never crossed, guaranteeing the invariance of the integral.

\section{\label{AppendixB}Proper time evaluation of Torque}

The Schwinger proper time formalism (Ref. \cite{Schwinger1951}) provides an alternative to the complex contour integration utilized in Section \ref{spatialaxion}, serving as a robust consistency check for the symmetry-breaking vacuum energy. For a purely spatial background gradient misaligned by an angle $\vartheta$ with respect to the plates' normal, the Euclidean background vector is $\mathbf{b}_E = (b \sin\vartheta, 0, b \cos\vartheta, 0)$. The differential operator governing the photon modes is $\mathcal{P}(\mathbf{k}_E) = \mathbf{k}_E^4 + b^2 \mathbf{k}_E^2 - (\mathbf{b}_E \cdot \mathbf{k}_E)^2$. The one-loop effective action is evaluated using the proper time representation of the logarithm:
\begin{eqnarray} \label{eq:B1}
\ln \mathcal{P}(\mathbf{k}_E) &=& - \int_0^\infty \frac{\mathrm{d}s}{s} \mathrm{e}^{-s \mathbf{k}_E^2} \nonumber \left[ \mathrm{e}^{-\mathbf{i}s X} + \mathrm{e}^{\mathbf{i} s X} \right],
\end{eqnarray}
where $X = \sqrt{b^2 \mathbf{k}_E^2 - (\mathbf{b}_E \cdot \mathbf{k}_E)^2}$. Expanding the resultant cosine to leading order in the background field yields:
\begin{eqnarray} \label{eq:B2}
\ln \mathcal{P}(\mathbf{k}_E) \approx - 2 \int_0^\infty \frac{\mathrm{d}s}{s} \mathrm{e}^{-s \mathbf{k}_E^2} + \int_0^\infty \mathrm{d}s  \;\mathrm{e}^{-s \mathbf{k}_E^2} s X^2 .
\end{eqnarray}
The $\mathcal{O}(b^0)$ term reproduces $\mathcal{E}^{(0)}_\text{Casimir} = -\pi^2 / 720 L_z^4$. The $\mathcal{O}(b^2)$ shift responsible for the torque is extracted by applying the Poisson summation formula $Q_m = 2mL_z$, expanding the momentum bracket, dropping the odd-parity cross-terms and then evaluating the Gaussian integrals.
\begin{equation} \label{eq:B3}
\mathcal{E}^{(2)}_\text{Casimir} = \sum_{m=1}^\infty \frac{b^2}{\pi^2 Q_m^2} \left( \frac{3}{8} - \frac{\sin^2\vartheta}{4} \right).
\end{equation}
Taking the angular gradient $\tau = -\partial \mathcal{E} / \partial \vartheta$ reproduces the torque derived via the dispersion relation.

To extend this formalism to the spatio-temporal background, we generalize the Euclidean vector to $\mathbf{b}_E = (b\sin\vartheta, 0, b\cos\vartheta, b_4)$. The temporal mixing emerges at fourth order in the proper time expansion:
\begin{equation} \label{eq:B4}
\ln \mathcal{P}(\mathbf{k}_E) \supset -2 \int_0^\infty \frac{\mathrm{d}s}{s} \mathrm{e}^{-s \mathbf{k}_E^2} \left( \frac{s^4}{24} X^4 \right) ,
\end{equation}
where $X^2 = \mathbf{b}_E^2 \mathbf{k}_E^2 - (\mathbf{b}_E \cdot \mathbf{k}_E)^2$. We decompose $X^2 = S + T$ into a purely spatial component $S = b^2 \mathbf{k}_E^2 - (\vec{b}\cdot\vec{k})^2$ and a temporal-mixing component $T = b_4^2 \vec{k}^2 - 2b_4 k_4 (\vec{b}\cdot\vec{k})$. Expanding $X^4 = (S+T)^2$, terms with odd powers of $k_4$ vanish under symmetric integration over the real line. The parity-even elements proportional to both $b_4^2$ and the spatial angle $\vartheta$ are isolated as: $X^4_\vartheta = 2b_4^2 (\vec{b}\cdot\vec{k})^2 (2k_4^2 - \vec{k}^2)$. Substituting $(\vec{b}\cdot\vec{k})^2 = b^2(k_1 \sin\vartheta + k_3 \cos\vartheta)^2$ and discarding the transverse $k_1 k_3$ cross-term (which integrates to zero), the angular derivative yields the effective torque operator:
\begin{equation} \label{eq:B5}
\frac{\partial X^4_\vartheta}{\partial \vartheta} = 2b_4^2 b^2 \sin(2\vartheta) (k_1^2 - k_3^2)(2k_4^2 - k_1^2 - k_2^2 - k_3^2).
\end{equation}
The $\mathcal{O}(b_4^2 b^2)$ torque contribution is then given by:
\begin{equation} \label{eq:B6}
\tau^{(4)} = \frac{1}{12} \sum_{m=1}^\infty \int \frac{\mathrm{d}^4 k_E}{(2\pi)^4} \int_0^\infty \mathrm{d}s \, s^3 \mathrm{e}^{-s \mathbf{k}_E^2} \left( \frac{\partial X^4_\vartheta}{\partial \vartheta} \right) \mathrm{e}^{\mathbf{i} Q_m k_3}.
\end{equation}
The $\mathcal{O}(b_4^2 b^2)$ torque contribution is evaluated by replacing the transverse and temporal momenta with their standard Gaussian expectations, $\langle k_i^2 \rangle = 1/(2s)$. The remaining longitudinal $k_3$ integration is performed by differentiating the generating function $J_0 = \sqrt{\pi/s} \, \mathrm{e}^{-Q_m^2/4s}$ recursively with respect to the proper time $s$. Integrating out $s$ evaluates the momentum brackets to a mode-independent scalar:
\begin{align} \label{eq:B7}
\tau^{(4)} = \frac{b_4^2 b^2 \sin(2\vartheta)}{96 \pi^2} \sum_{m=1}^\infty \int_0^\infty \mathrm{d}s \left( -\frac{5Q_m^2}{8s^2} + \frac{Q_m^4}{16s^3} \right) \mathrm{e}^{-Q_m^2/4s}.
\end{align}
Applying zeta function regularization and analytically continuing to Minkowski space via $b_4^2 = -b_0^2$, we reproduce the topological spatio-temporal torque.

\section{\label{AppendixC} Validity of the spatial force expression}

This is a check on whether the spatial force (per unit area) in Eq. \eqref{eq:17} in this work reduces to the closed-form force per unit area derived by Ref. \cite{Fukushima2019} in the aligned limit $\vartheta \to 0$.
Under the limit $\vartheta \to 0$, $\tilde{A} = u/Q_m$, $\tilde{C} = u/Q_m^2$ and $\lambda = Q_m$. Now the force per unit area is given by $F/A = - \mathcal{E} + L_z \mathcal{F}$ where:
\begin{equation} \label{eq:C1}
\mathcal{F} = -\frac{2}{\pi^2} \sum_{m=1}^{\infty} \frac{m}{Q_m} \int_0^{\infty} \mathrm{d}u \, u^2 \operatorname{Re} \left[ \mathrm{e}^{-Q_m R_+} \left( R_+ + \frac{1}{Q_m} \right) \right],
\end{equation} and the energy density becomes:
\begin{equation} \label{eq:C2}
\mathcal{E} = -\frac{1}{\pi^2} \sum_{m=1}^{\infty} \frac{1}{Q_m} \int_0^{\infty} \mathrm{d}u \, u^2 \operatorname{Re} \left[ \mathrm{e}^{-Q_m R_+ }\right].
\end{equation}
We analyze $R_+$ here. Because $\left(\sqrt{u^2 + \mathbf{i}bu}\right)^* = \sqrt{(-u)^2 + \mathbf{i}b(-u)}$, the integrand satisfies $f(-u) = f^*(u)$. This shift can be validated by the same procedure as in Appendix \ref{AppendixA}. We can rewrite the half-line real part as a symmetric integral over the entire real line. Complete the square under the radical via $u^2 + \mathbf{i}bu = (u + \mathbf{i}\frac{b}{2})^2 + (\frac{b}{2})^2$. Shift the integration variable $v = u + \mathbf{i}\frac{b}{2}$ and let $a = \frac{b}{2}$. Expanding $(v - \mathbf{i}a)^2 = v^2 - \mathbf{i}2av - a^2$, the odd term $-\mathbf{i}2av$ vanishes under symmetric integration. Folding the even domain back to the half-line gives purely real integrals:
\begin{align}
I_E &\equiv \int_0^\infty \mathrm{d}v \, (v^2 - a^2) \mathrm{e}^{-Q_m \sqrt{v^2 + a^2}} \text{, and} \label{eq:C3}\\
I_F &\equiv \frac{1}{Q_m} \int_0^\infty \mathrm{d}v \, (v^2 - a^2) \mathrm{e}^{-Q_m \sqrt{v^2 + a^2}} \left( \sqrt{v^2 + a^2} + \frac{1}{Q_m} \right).\label{eq:C4}
\end{align}
Substitute $v = a \sinh t$, apply $\cosh t (\sinh^2 t - 1) = \frac{1}{4}\cosh(3t) - \frac{5}{4}\cosh t$, and utilize the recurrence $K_{\nu+1}(z) = K_{\nu-1}(z) + \frac{2\nu}{z} K_\nu(z)$ with $z = Q_m a = mbL_z$ to evaluate the integrals. Here $K_\nu(z) = \int_0^\infty \mathrm{e}^{-z \cosh t} \cosh(\nu t) \mathrm{d}t$ are modified Bessel functions of the second kind.
\begin{align}
I_E &= a^3 \left[ -\frac{z}{2} K_2(z) + \frac{z}{2} K_0(z) + \frac{1}{z} K_2(z) \right] ,\label{eq:C5}\\
I_F &= a^5 \left[ \frac{4}{z^3}K_2(z) - \frac{1}{2z}K_2(z) - \frac{1}{2z}K_0(z) \right].\label{eq:C6}
\end{align}
Substituting $I_E$ and $I_F$ back into the density summations produces:
\begin{align}
\mathcal{E} &= \frac{b^4}{16\pi^2} \sum_{m=1}^\infty \left[ \frac{1}{2} K_2(z) - \frac{1}{2} K_0(z) - \frac{1}{z^2} K_2(z) \right] ,\label{eq:C7}\\
L_z \mathcal{F} &= -\frac{b^4}{16\pi^2} \sum_{m=1}^\infty \left[ \frac{4}{z^2} K_2(z) - \frac{1}{2} K_2(z) - \frac{1}{2} K_0(z) \right].\label{eq:C8}
\end{align}
Using these expressions, we get:
\begin{equation} \label{eq:C9}
F/A = -\frac{b^4}{16\pi^2} \sum_{m=1}^\infty \left[ \frac{3K_2(mbL_z)}{(mbL_z)^2} - K_0(mbL_z) \right],
\end{equation}
which agrees with the results in Ref. \cite{Fukushima2019}.

%\nocite{*}
\bibliography{apssamp}

@PREAMBLE{
 "\providecommand{\noopsort}[1]{}" 
 # "\providecommand{\singleletter}[1]{#1}%" 
}

@article{Casimir1948,
    author = "Casimir, H. B. G.",
    title = "{On the attraction between two perfectly conducting plates}",
    journal = "Indag. Math.",
    volume = "10",
    number = "4",
    pages = "261--263",
    year = "1948"
}

@article{Fukushima2019,
  title = {Anomalous Casimir effect in axion electrodynamics},
  author = {Fukushima, Kenji and Imaki, Shota and Qiu, Zebin},
  journal = {Phys. Rev. D},
  volume = {100},
  issue = {4},
  pages = {045013},
  numpages = {8},
  year = {2019},
  month = {Aug},
  publisher = {American Physical Society},
  doi = {10.1103/PhysRevD.100.045013},
  url = {https://link.aps.org/doi/10.1103/PhysRevD.100.045013}
}

@article{Ema2023,
  title = {Zero Casimir force in axion electrodynamics and the search for a new force},
  author = {Ema, Yohei and Hazumi, Masashi and Iizuka, Hideo and Mukaida, Kyohei and Nakayama, Kazunori},
  journal = {Phys. Rev. D},
  volume = {108},
  issue = {1},
  pages = {016009},
  numpages = {16},
  year = {2023},
  month = {Jul},
  publisher = {American Physical Society},
  doi = {10.1103/PhysRevD.108.016009},
  url = {https://link.aps.org/doi/10.1103/PhysRevD.108.016009}
}

@article{Favitta2023,
   title={Axion electrodynamics: Green’s functions, zero-point energy and optical activity},
   volume={455},
   ISSN={0003-4916},
   url={http://dx.doi.org/10.1016/j.aop.2023.169396},
   DOI={10.1016/j.aop.2023.169396},
   journal={Annals of Physics},
   publisher={Elsevier BV},
   author={Favitta, Amedeo M. and Brevik, Iver H. and Chaichian, Moshe M.},
   year={2023},
   month=Aug, pages={169396} }

@article{PecceiQuinn1977,
  title = {$\mathrm{CP}$ Conservation in the Presence of Pseudoparticles},
  author = {Peccei, R. D. and Quinn, Helen R.},
  journal = {Phys. Rev. Lett.},
  volume = {38},
  issue = {25},
  pages = {1440--1443},
  numpages = {0},
  year = {1977},
  month = {Jun},
  publisher = {American Physical Society},
  doi = {10.1103/PhysRevLett.38.1440},
  url = {https://link.aps.org/doi/10.1103/PhysRevLett.38.1440}
}

@article{Weinberg1978,
  title = {A New Light Boson?},
  author = {Weinberg, Steven},
  journal = {Phys. Rev. Lett.},
  volume = {40},
  issue = {4},
  pages = {223--226},
  numpages = {0},
  year = {1978},
  month = {Jan},
  publisher = {American Physical Society},
  doi = {10.1103/PhysRevLett.40.223},
  url = {https://link.aps.org/doi/10.1103/PhysRevLett.40.223}
}

@article{Wilczek1978,
  title = {Problem of Strong $P$ and $T$ Invariance in the Presence of Instantons},
  author = {Wilczek, F.},
  journal = {Phys. Rev. Lett.},
  volume = {40},
  issue = {5},
  pages = {279--282},
  numpages = {0},
  year = {1978},
  month = {Jan},
  publisher = {American Physical Society},
  doi = {10.1103/PhysRevLett.40.279},
  url = {https://link.aps.org/doi/10.1103/PhysRevLett.40.279}
}

@article{SikiviePRL1983,
  title = {Experimental Tests of the "Invisible" Axion},
  author = {Sikivie, P.},
  journal = {Phys. Rev. Lett.},
  volume = {51},
  issue = {16},
  pages = {1415--1417},
  numpages = {0},
  year = {1983},
  month = {Oct},
  publisher = {American Physical Society},
  doi = {10.1103/PhysRevLett.51.1415},
  url = {https://link.aps.org/doi/10.1103/PhysRevLett.51.1415}
}

@article{Wilczek1987,
  title = {Two applications of axion electrodynamics},
  author = {Wilczek, Frank},
  journal = {Phys. Rev. Lett.},
  volume = {58},
  issue = {18},
  pages = {1799--1802},
  numpages = {0},
  year = {1987},
  month = {May},
  publisher = {American Physical Society},
  doi = {10.1103/PhysRevLett.58.1799},
  url = {https://link.aps.org/doi/10.1103/PhysRevLett.58.1799}
}

@article{Brevik2021,
   title={Axion Electrodynamics and the Axionic Casimir Effect},
   volume={7},
   ISSN={2218-1997},
   url={http://dx.doi.org/10.3390/universe7050133},
   DOI={10.3390/universe7050133},
   number={5},
   journal={Universe},
   publisher={MDPI AG},
   author={Brevik, Iver},
   year={2021},
   month=May, pages={133} }

@article{Somers2018,
  author     = {Somers, David A. T. and Garrett, Joseph L. and Palm, Kevin J. and Munday, Jeremy N.},
  title      = {Measurement of the Casimir torque},
  journal    = {Nature},
  volume     = {564},
  number     = {7736},
  pages      = {386--389},
  year       = {2018},
  publisher = {Nature Publishing Group},
  doi        = {10.1038/s41586-018-0777-8}
}

@misc{Iannuzzi2003,
      title={New Challenges and Directions in Casimir Force Experiments}, 
      author={Davide Iannuzzi and Ian Gelfand and Mariangela Lisanti and Federico Capasso},
      year={2003},
      eprint={quant-ph/0312043},
      archivePrefix={arXiv},
      primaryClass={quant-ph},
      url={https://arxiv.org/abs/quant-ph/0312043}, 
}

@article{
Kucukoz2024,
author = {Betül Küçüköz  and Oleg V. Kotov  and Adriana Canales  and Alexander Yu. Polyakov  and Abhay V. Agrawal  and Tomasz J. Antosiewicz  and Timur O. Shegai },
title = {Quantum trapping and rotational self-alignment in triangular Casimir microcavities},
journal = {Science Advances},
volume = {10},
number = {17},
pages = {eadn1825},
year = {2024},
doi = {10.1126/sciadv.adn1825},
URL = {https://www.science.org/doi/abs/10.1126/sciadv.adn1825},
eprint = {https://www.science.org/doi/pdf/10.1126/sciadv.adn1825}}

@misc{zhang2024Comment,
      title={Comment on "Quantum trapping and rotational self-alignment in triangular Casimir microcavities"}, 
      author={Zhentao Zhang},
      year={2024},
      eprint={2406.17742},
      archivePrefix={arXiv},
      primaryClass={cond-mat.mes-hall},
      url={https://arxiv.org/abs/2406.17742}, 
}

@article{Guerout2015,
   title={Casimir torque between nanostructured plates},
   volume={111},
   ISSN={1286-4854},
   url={http://dx.doi.org/10.1209/0295-5075/111/44001},
   DOI={10.1209/0295-5075/111/44001},
   number={4},
   journal={EPL (Europhysics Letters)},
   publisher={IOP Publishing},
   author={Guérout, R. and Genet, C. and Lambrecht, A. and Reynaud, S.},
   year={2015},
   month=Aug, pages={44001} }

@article{XuLi2017,
   title={Detecting Casimir torque with an optically levitated nanorod},
   volume={96},
   ISSN={2469-9934},
   url={http://dx.doi.org/10.1103/PhysRevA.96.033843},
   DOI={10.1103/physreva.96.033843},
   number={3},
   journal={Physical Review A},
   publisher={American Physical Society (APS)},
   author={Xu, Zhujing and Li, Tongcang},
   year={2017},
   month={9} }

@article{Spreng2022,
author = {Spreng, Benjamin and Gong, Tao and Munday, Jeremy N.},
title = {Recent developments on the Casimir torque},
journal = {International Journal of Modern Physics A},
volume = {37},
number = {19},
pages = {2241011},
year = {2022},
doi = {10.1142/S0217751X22410111},

URL = {   
        https://doi.org/10.1142/S0217751X22410111
},
eprint = {     
        https://doi.org/10.1142/S0217751X22410111
}
}

@article{Hu2025,
  title = {Casimir radiation with Weyl semimetals},
  author = {Hu, Yang and Wu, Xiaohu and Liu, Haotuo and Antezza, Mauro and Huang, Xiuquan},
  journal = {Phys. Rev. Res.},
  volume = {7},
  issue = {4},
  pages = {043328},
  numpages = {7},
  year = {2025},
  month = {Dec},
  publisher = {American Physical Society},
  doi = {10.1103/ldyf-753c},
  url = {https://link.aps.org/doi/10.1103/ldyf-753c}
}

@article{Ahn2019,
    author = "Ahn, Jonghoon and Xu, Zhujing and Bang, Jaehoon and Ju, Peng and Gao, Xingyu and Li, Tongcang",
    title = "{Ultrasensitive torque detection with an optically levitated nanorotor}",
    eprint = "1908.03453",
    archivePrefix = "arXiv",
    primaryClass = "physics.app-ph",
    doi = "10.1038/s41565-019-0605-9",
    journal = "Nature Nanotech.",
    volume = "15",
    pages = "89",
    year = "2020"
}

@article{Ahn2018,
  title = {Optically Levitated Nanodumbbell Torsion Balance and GHz Nanomechanical Rotor},
  author = {Ahn, Jonghoon and Xu, Zhujing and Bang, Jaehoon and Deng, Yu-Hao and Hoang, Thai M. and Han, Qinkai and Ma, Ren-Min and Li, Tongcang},
  journal = {Phys. Rev. Lett.},
  volume = {121},
  issue = {3},
  pages = {033603},
  numpages = {5},
  year = {2018},
  month = {Jul},
  publisher = {American Physical Society},
  doi = {10.1103/PhysRevLett.121.033603},
  url = {https://link.aps.org/doi/10.1103/PhysRevLett.121.033603}
}

@article{GrahamRajendran2011,
  title = {Axion dark matter detection with cold molecules},
  author = {Graham, Peter W. and Rajendran, Surjeet},
  journal = {Phys. Rev. D},
  volume = {84},
  issue = {5},
  pages = {055013},
  numpages = {13},
  year = {2011},
  month = {Sep},
  publisher = {American Physical Society},
  doi = {10.1103/PhysRevD.84.055013},
  url = {https://link.aps.org/doi/10.1103/PhysRevD.84.055013}
}

@article{Evans2019,
  title = {Refinement of the standard halo model for dark matter searches in light of the Gaia Sausage},
  author = {Evans, N. Wyn and O'Hare, Ciaran A. J. and McCabe, Christopher},
  journal = {Phys. Rev. D},
  volume = {99},
  issue = {2},
  pages = {023012},
  numpages = {18},
  year = {2019},
  month = {Jan},
  publisher = {American Physical Society},
  doi = {10.1103/PhysRevD.99.023012},
  url = {https://link.aps.org/doi/10.1103/PhysRevD.99.023012}
}

@article{GrahamIrastorza2015,
   title={Experimental Searches for the Axion and Axion-Like Particles},
   volume={65},
   ISSN={1545-4134},
   url={http://dx.doi.org/10.1146/annurev-nucl-102014-022120},
   DOI={10.1146/annurev-nucl-102014-022120},
   number={1},
   journal={Annual Review of Nuclear and Particle Science},
   publisher={Annual Reviews},
   author={Graham, Peter W. and Irastorza, Igor G. and Lamoreaux, Steven K. and Lindner, Axel and van Bibber, Karl A.},
   year={2015},
   month=Oct, pages={485–514} }

@article{ZyuzinBurkov2012,
  title = {Topological response in Weyl semimetals and the chiral anomaly},
  author = {Zyuzin, A. A. and Burkov, A. A.},
  journal = {Phys. Rev. B},
  volume = {86},
  issue = {11},
  pages = {115133},
  numpages = {8},
  year = {2012},
  month = {Sep},
  publisher = {American Physical Society},
  doi = {10.1103/PhysRevB.86.115133},
  url = {https://link.aps.org/doi/10.1103/PhysRevB.86.115133}
}

@article{ChenWuBurkov2013,
  title = {Axion response in Weyl semimetals},
  author = {Chen, Y. and Wu, Si and Burkov, A. A.},
  journal = {Phys. Rev. B},
  volume = {88},
  issue = {12},
  pages = {125105},
  numpages = {6},
  year = {2013},
  month = {Sep},
  publisher = {American Physical Society},
  doi = {10.1103/PhysRevB.88.125105},
  url = {https://link.aps.org/doi/10.1103/PhysRevB.88.125105}
}

@article{Schwinger1951,
  author  = "Schwinger, Julian",
  title   = "{On Gauge Invariance and Vacuum Polarization}",
  journal = "Physical Review",
  volume  = "82",
  number  = "5",
  pages   = "664--679",
  year    = "1951",
  doi     = "10.1103/PhysRev.82.664",
  url     = "https://link.aps.org/doi/10.1103/PhysRev.82.664"
}

@article{KlimchitskayaMostepanenkoMohideen2009,
   title={The Casimir force between real materials: Experiment and theory},
   volume={81},
   ISSN={1539-0756},
   url={http://dx.doi.org/10.1103/RevModPhys.81.1827},
   DOI={10.1103/revmodphys.81.1827},
   number={4},
   journal={Reviews of Modern Physics},
   publisher={American Physical Society (APS)},
   author={Klimchitskaya, G. L. and Mohideen, U. and Mostepanenko, V. M.},
   year={2009},
   month=Dec, pages={1827–1885} }

@article{ParsegianWeiss1972,
  author  = {Parsegian, V. A. and Weiss, G. H.},
  title   = {Dielectric anisotropy and the van der {W}aals interaction between bulk media},
  journal = {J. Adhes.},
  volume  = {3},
  number  = {4},
  pages   = {259--267},
  year    = {1972},
  doi     = {10.1080/00218467208072197}
}

@article{Barash1978,
  author    = {Barash, Yu. S.},
  title     = {Moment of van der Waals forces between anisotropic bodies},
  journal   = {Radiophysics and Quantum Electronics},
  year      = {1978},
  volume    = {21},
  number    = {11},
  pages     = {1138--1143},
  doi       = {10.1007/BF02121382},
  url       = {https://doi.org/10.1007/BF02121382}
}

@ARTICLE{vanEnk1995,
       author = {{van Enk}, S.~J.},
        title = "{Casimir torque between dielectrics}",
      journal = {\pra},
         year = 1995,
        month = oct,
       volume = {52},
       number = {4},
        pages = {2569-2575},
          doi = {10.1103/PhysRevA.52.2569},
       adsurl = {https://ui.adsabs.harvard.edu/abs/1995PhRvA..52.2569V}
}

@article{Munday2005,
  title = {Torque on birefringent plates induced by quantum fluctuations},
  author = {Munday, Jeremy N. and Iannuzzi, Davide and Barash, Yuri and Capasso, Federico},
  journal = {Phys. Rev. A},
  volume = {71},
  issue = {4},
  pages = {042102},
  numpages = {9},
  year = {2005},
  month = {Apr},
  publisher = {American Physical Society},
  doi = {10.1103/PhysRevA.71.042102},
  url = {https://link.aps.org/doi/10.1103/PhysRevA.71.042102}
}

@article{Munday2008Erratum,
  title = {Erratum: Torque on birefrigent plates induced by quantum fluctuations [Phys. Rev. A 71, 042102 (2005)]},
  author = {Munday, Jeremy N. and Iannuzzi, Davide and Barash, Yuri and Capasso, Federico},
  journal = {Phys. Rev. A},
  volume = {78},
  issue = {2},
  pages = {029906},
  numpages = {1},
  year = {2008},
  month = {8},
  publisher = {American Physical Society},
  doi = {10.1103/PhysRevA.78.029906},
  url = {https://link.aps.org/doi/10.1103/PhysRevA.78.029906}
}

@article{CarrollFieldJackiw1990,
  title = {Limits on a Lorentz- and parity-violating modification of electrodynamics},
  author = {Carroll, Sean M. and Field, George B. and Jackiw, Roman},
  journal = {Phys. Rev. D},
  volume = {41},
  issue = {4},
  pages = {1231--1240},
  numpages = {0},
  year = {1990},
  month = {2},
  publisher = {American Physical Society},
  doi = {10.1103/PhysRevD.41.1231},
  url = {https://link.aps.org/doi/10.1103/PhysRevD.41.1231}
}

@article{QiHughesZhang2008,
  title = {Topological field theory of time-reversal invariant insulators},
  author = {Qi, Xiao-Liang and Hughes, Taylor L. and Zhang, Shou-Cheng},
  journal = {Phys. Rev. B},
  volume = {78},
  issue = {19},
  pages = {195424},
  numpages = {43},
  year = {2008},
  month = {Nov},
  publisher = {American Physical Society},
  doi = {10.1103/PhysRevB.78.195424},
  url = {https://link.aps.org/doi/10.1103/PhysRevB.78.195424}
}

@article{EssinMooreVanderbilt2009,
  title = {Magnetoelectric Polarizability and Axion Electrodynamics in Crystalline Insulators},
  author = {Essin, Andrew M. and Moore, Joel E. and Vanderbilt, David},
  journal = {Phys. Rev. Lett.},
  volume = {102},
  issue = {14},
  pages = {146805},
  numpages = {4},
  year = {2009},
  month = {Apr},
  publisher = {American Physical Society},
  doi = {10.1103/PhysRevLett.102.146805},
  url = {https://link.aps.org/doi/10.1103/PhysRevLett.102.146805}
}

@article{WanTurnerVishwanathSavrasov2011,
  title = {Topological semimetal and Fermi-arc surface states in the electronic structure of pyrochlore iridates},
  author = {Wan, Xiangang and Turner, Ari M. and Vishwanath, Ashvin and Savrasov, Sergey Y.},
  journal = {Phys. Rev. B},
  volume = {83},
  issue = {20},
  pages = {205101},
  numpages = {9},
  year = {2011},
  month = {May},
  publisher = {American Physical Society},
  doi = {10.1103/PhysRevB.83.205101},
  url = {https://link.aps.org/doi/10.1103/PhysRevB.83.205101}
}

@article{ArmitageMeleVishwanath2018,
  title = {Weyl and Dirac semimetals in three-dimensional solids},
  author = {Armitage, N. P. and Mele, E. J. and Vishwanath, Ashvin},
  journal = {Rev. Mod. Phys.},
  volume = {90},
  issue = {1},
  pages = {015001},
  numpages = {57},
  year = {2018},
  month = {Jan},
  publisher = {American Physical Society},
  doi = {10.1103/RevModPhys.90.015001},
  url = {https://link.aps.org/doi/10.1103/RevModPhys.90.015001}
}

@article{KennethKlich2006,
  title = {Opposites Attract: A Theorem about the Casimir Force},
  author = {Kenneth, O. and Klich, I.},
  journal = {Phys. Rev. Lett.},
  volume = {97},
  pages = {160401},
  year = {2006},
  doi = {10.1103/PhysRevLett.97.160401}
}

@article{PreskillWiseWilczek1983,
    author = "Preskill, John and Wise, Mark B. and Wilczek, Frank",
    editor = "Srednicki, M. A.",
    title = "{Cosmology of the Invisible Axion}",
    reportNumber = "HUTP-82-A048, NSF-ITP-82-103",
    doi = "10.1016/0370-2693(83)90637-8",
    journal = "Phys. Lett. B",
    volume = "120",
    pages = "127--132",
    year = "1983"
}

@article{AbbottSikivie1983,
title = {A cosmological bound on the invisible axion},
journal = {Physics Letters B},
volume = {120},
number = {1},
pages = {133-136},
year = {1983},
issn = {0370-2693},
doi = {https://doi.org/10.1016/0370-2693(83)90638-X},
url = {https://www.sciencedirect.com/science/article/pii/037026938390638X},
author = {L.F. Abbott and P. Sikivie}
}

@article{DineFischler1983,
    author = "Dine, Michael and Fischler, Willy",
    editor = "Srednicki, M. A.",
    title = "{The Not So Harmless Axion}",
    reportNumber = "UPR-0201T",
    doi = "10.1016/0370-2693(83)90639-1",
    journal = "Phys. Lett. B",
    volume = "120",
    pages = "137--141",
    year = "1983"
}

@article{JiangWilczek2019,
  title = {Chiral Casimir forces: Repulsive, enhanced, tunable},
  author = {Jiang, Qing-Dong and Wilczek, Frank},
  journal = {Phys. Rev. B},
  volume = {99},
  issue = {12},
  pages = {125403},
  numpages = {10},
  year = {2019},
  month = {Mar},
  publisher = {American Physical Society},
  doi = {10.1103/PhysRevB.99.125403},
  url = {https://link.aps.org/doi/10.1103/PhysRevB.99.125403}
}

@article{KharlanovZhukovsky2010,
  title = {Casimir effect within $D=3+1$ Maxwell-Chern-Simons electrodynamics},
  author = {Kharlanov, O. G. and Zhukovsky, V. Ch.},
  journal = {Phys. Rev. D},
  volume = {81},
  issue = {2},
  pages = {025015},
  numpages = {18},
  year = {2010},
  month = {Jan},
  publisher = {American Physical Society},
  doi = {10.1103/PhysRevD.81.025015},
  url = {https://link.aps.org/doi/10.1103/PhysRevD.81.025015}
}

@article{GrushinCortijo2011,
  title = {Tunable Casimir Repulsion with Three-Dimensional Topological Insulators},
  author = {Grushin, Adolfo G. and Cortijo, Alberto},
  journal = {Phys. Rev. Lett.},
  volume = {106},
  issue = {2},
  pages = {020403},
  numpages = {4},
  year = {2011},
  month = {Jan},
  publisher = {American Physical Society},
  doi = {10.1103/PhysRevLett.106.020403},
  url = {https://link.aps.org/doi/10.1103/PhysRevLett.106.020403}
}

@article{RodriguezLopezGrushin2014,
  title = {Repulsive Casimir Effect with Chern Insulators},
  author = {Rodriguez-Lopez, Pablo and Grushin, Adolfo G.},
  journal = {Phys. Rev. Lett.},
  volume = {112},
  issue = {5},
  pages = {056804},
  numpages = {5},
  year = {2014},
  month = {Feb},
  publisher = {American Physical Society},
  doi = {10.1103/PhysRevLett.112.056804},
  url = {https://link.aps.org/doi/10.1103/PhysRevLett.112.056804}
}

@article{TseMacDonald2012,
  title = {Quantized Casimir Force},
  author = {Tse, Wang-Kong and MacDonald, A. H.},
  journal = {Phys. Rev. Lett.},
  volume = {109},
  issue = {23},
  pages = {236806},
  numpages = {5},
  year = {2012},
  month = {Dec},
  publisher = {American Physical Society},
  doi = {10.1103/PhysRevLett.109.236806},
  url = {https://link.aps.org/doi/10.1103/PhysRevLett.109.236806}
}

@article{CaoZhitnitsky2017,
   title={Axion detection via topological Casimir effect},
   volume={96},
   ISSN={2470-0029},
   url={http://dx.doi.org/10.1103/PhysRevD.96.015013},
   DOI={10.1103/physrevd.96.015013},
   number={1},
   journal={Physical Review D},
   publisher={American Physical Society (APS)},
   author={Cao, ChunJun and Zhitnitsky, Ariel},
   year={2017},
   month={7} }

@article{Lifshitz1956,
  author  = {Lifshitz, E. M.},
  title   = {The theory of molecular attractive forces between solids},
  journal = {Sov. Phys. JETP},
  volume  = {2},
  pages   = {73--83},
  year    = {1956},
  url={https://jetp.ras.ru/cgi-bin/dn/e_002_01_0073.pdf},
}

@article{WilsonAlloccaGalitski2015,
  title = {Repulsive Casimir force between Weyl semimetals},
  author = {Wilson, Justin H. and Allocca, Andrew A. and Galitski, Victor},
  journal = {Phys. Rev. B},
  volume = {91},
  issue = {23},
  pages = {235115},
  numpages = {5},
  year = {2015},
  month = {6},
  publisher = {American Physical Society},
  doi = {10.1103/PhysRevB.91.235115},
  url = {https://link.aps.org/doi/10.1103/PhysRevB.91.235115}
}
\end{document}